%% file: main.tex
\documentclass[conference]{IEEEtran}
\usepackage[usenames,dvipsnames,svgnames,table]{xcolor}

\ifCLASSINFOpdf
  \usepackage[pdftex]{graphicx}
\fi
\usepackage{amsmath}
\usepackage{algorithmic}
\usepackage{booktabs}
\usepackage[skip=2pt]{caption}
\usepackage{amssymb} 
\usepackage{subcaption}
\usepackage{multirow}
\usepackage[tikz]{mdframed}
\usepackage{xcolor}
\usepackage{tikz}
\usepackage{hyperref}
\hypersetup{
    colorlinks=true,
    linkcolor=black,
    citecolor=black,
}
\input{macros}

\begin{document}
\title{Attacks and Mitigations for Distributed Governance of Agentic AI under Byzantine Adversaries}

\author{\IEEEauthorblockN{Matthew D. Laws}
	\IEEEauthorblockA{Northeastern University\\
		laws.ma@northeastern.edu}
	\and
	\IEEEauthorblockN{Alina Oprea}
	\IEEEauthorblockA{Northeastern University\\
		a.oprea@northeastern.edu}
    \and
	\IEEEauthorblockN{Cristina Nita-Rotaru}
	\IEEEauthorblockA{Northeastern University\\
		c.nitarotaru@northeastern.edu}}

\maketitle

\input{sections/0-abstract}

\IEEEpeerreviewmaketitle

\input{sections/1-introduction}

\input{sections/2-background}
\input{sections/3-attacks}

\input{sections/4-bft}

\input{sections/5-monitor_and_audit}
\input{sections/6-analysis}

\input{sections/7-hybrid}

\input{sections/8-experiments}
\input{sections/9-related_work}
\input{sections/10-conclusion}

\bibliographystyle{IEEEtran}
\bibliography{bib/references}
\input{sections/A-appendix}

\end{document}

%% file: macros.tex
\newcommand{\provider}{\texttt{Provider}}

\newcommand{\saga}{SAGA}
\newcommand{\controller}{controller}

\newcommand{\sagabft}{SAGA-BFT}
\newcommand{\sagaaud}{SAGA-AUD}
\newcommand{\sagamon}{SAGA-MON}
\newcommand{\sagahyb}{SAGA-HYB}

\newcommand{\Figure}[1]{Figure~\ref{#1}}

\newcommand{\attack}[2]{\textit{Attack #1}~(#2).\quad}

%% file: sections/0-abstract.tex
\begin{abstract}
Agentic AI governance is a critical component of agentic AI infrastructure ensuring that agents follow their owner’s communication and interaction policies, and providing protection against attacks from malicious agents. 
The state-of-the-art solution, SAGA, assumes a logically centralized point of trust, the \provider, which serves as a repository for user and agent information and actively enforces policies. While \saga~provides protection against malicious agents, it remains vulnerable to a malicious \provider~that deviates from the protocol, undermining the security of the identity and access control infrastructure. Deployment on both private and public clouds, each susceptible to insider threats, further increases the risk of \provider~compromise.

In this work, we analyze the attacks that can be mounted from a compromised \provider, taking into account the different system components and realistic deployments. We identify and execute several concrete attacks with devastating effects: undermining agent attributability, extracting private data, or bypassing access control.
We then present three types of solutions for securing the \provider~that offer different trade-offs between security and performance.
We first present \sagabft, a fully byzantine-resilient architecture that provides the strongest protection, but incurs significant performance degradation, due to the high-cost of byzantine resilient protocols. We then propose \sagamon~and \sagaaud, two novel solutions that leverage lightweight server-side monitoring or client-side auditing to provide protection against most classes of attacks with minimal overhead. Finally, we propose \sagahyb, a hybrid architecture that combines byzantine-resilience with monitoring and auditing to trade-off security for performance. We evaluate all the architectures and compare them with \saga.  We discuss which solution is best and under what conditions.
\end{abstract}

%% file: sections/1-introduction.tex
\section{Introduction}
AI agents -- systems capable of adaptively pursuing complex goals in dynamic, real-world environments with minimal direct supervision~\cite{bellogin2025systemic} -- are rapidly moving from research prototypes to production deployments. Several frameworks 
(AutoGen~\cite{autogen}, MetaGPT~\cite{metagpt}) were introduced 
that enable developers to build autonomous agents atop large language models (LLMs) and a growing number of enterprises are integrating these agents into operational workflows. In addition, protocols like A2A \cite{google2025a2a} and MCP ~\cite{anthropic2024mcp} allow agents to communicate with each other or to call external tools. None of these come with in-built security and they primarily focus on interoperability and deployment.

The rapid deployment of agentic AI without appropriate secure infrastructure has left the current security stack vulnerable to attacks from untrusted or compromised agents. Recent incidents underscore the severity of this gap. In June 2025, Microsoft discovered CVE-2025-32711 (EchoLeak)~\cite{echoleak}, a zero-click indirect prompt injection in which a malicious email exfiltrates sensitive data from Copilot's context window without any user interaction. In July 2025, Replit's AI agent deleted an entire production database belonging to SaaStr despite explicit user instructions to the contrary~\cite{kahn2025replit}. And in November 2025, Anthropic disclosed the first documented AI-agent-orchestrated cyberattack~\cite{anthropic2025gtg1002}. These examples illustrate how deeply agents can be embedded in critical workflows and the risks they introduce. It is therefore essential to design secure governance architectures that uniquely identify agents, authenticate their interactions, ensure they follow their owner's policies, attribute their actions to the users that own them, and revoke malicious agents from the system~\cite{shavit2023practices}.

Several solutions have been proposed, including requirements for agent identities~\cite{chan_ids_2024}, capability taxonomies~\cite{muscariello2025agntcy}, attribution mechanisms~\cite{chan_infrastructure_2025}, authorization with delegation~\cite{south_authenticated_2025}, indexing~\cite{nanda_index_2025} and decentralized identifiers~\cite{huang2025fortifying}, but these remain largely theoretical, lacking implementation, empirical evaluation, or provable guarantees. More concrete proposals exist but each carries significant limitations: constitution-based defenses~\cite{hua2024trustagent} and just-in-time policy determination~\cite{tsai2025contextual} rely on the LLM itself for enforcement%
, while protocol-hardening approaches~\cite{louck2025improving, habler2025building, hou2026smcp} lack empirical evaluation.

One exception, is \saga~\cite{saga_ndss2026}, a publicly available \cite{saga_git} governance architecture for agentic AI systems that enables user-controlled agent management, enforces accountability through verifiable governance, and incorporates cryptographic mechanisms 
to provide forward secrecy and secure inter-agent communication. \saga~ relies on a trusted centralized \provider, that serves as a registry for user and agent information, and plays an active role in policy enforcement. 

While SAGA provides protection against malicious agents, it remains
vulnerable to a malicious \provider. Deployment on both private and public
clouds, each susceptible to insider threats, further increases the
risk of \provider~compromise. In cloud and multi-tenant infrastructure, the shared nature of underlying hardware exposes containers to side-channel attacks that can compromise co-located components \cite{ristenpart2009hey, jarkas2025container} and privilege escalation (e.g., CVE-2024-21626, CVE-2019-5736, CVE-2023-1260). 
A compromised \provider~could selectively not follow the protocols and undermine the safety guarantees that SAGA is designed to provide. Such behavior is typically referred to as byzantine behavior, and systems designed to withstand it are referred to as byzantine-resilient.

In this work, we analyze the attacks that can happen from a byzantine \provider, taking into account the different components of the system and realistic deployments. We demonstrate that such compromises can have a devastating effect by executing concrete attacks that allow an attacker to undermine agent attributability, extract private data, prevent authorized communication, or allow unauthorized one. 

Mitigating byzantine attacks is a challenging task, particularly for systems that have to meet high throughput and low latency requirements such as \saga's \provider, and where the access control nature of the application, requires deterministic finality. On one side byzantine state machine replication with deterministic finality is well understood and many solutions exist based on the pioneering BFT protocol \cite{pbft}. On the other side, such protocols are notoriously expensive as they require multiple rounds of communication between a significant set of participants (typically two thirds) and difficult to implement and deploy in practice. 

We present three types of solutions that offer different trade-offs between security and performance. We first present \sagabft, a fully byzantine-resilient architecture that provides the strongest protection. Specifically,
it prevents all the attacks from any compromised component of the \provider~(e.g. access control engine, database) by replicating these components on a set of replicas, binding their decisions cryptographically, and requiring a quorum of honest replicas for each operation. This architecture incurs significant performance degradation, due to high-cost of byzantine resilient protocols and we provide insights into this cost.

We then propose \sagamon~and \sagaaud, two novel solutions that rely on monitoring and auditing to overcome the cost of byzantine resilient replication.  \sagamon~can be run by the cloud operator or the agentic system operator to detect malicious \provider~behavior by analyzing network, host, and database logs. 
\sagaaud~is intended to be run by the agents owners, where a small set of auditing agents can perform probing requests to check the correctness of the  \provider's answers. Both approaches have small overhead,  can be performed with different levels of periodicity and intensity and we provide a security analysis that shows how to configure them to ensure high-probability of detection.

Finally, we propose \sagahyb, a hybrid architecture that combines byzantine-resilience with monitoring and auditing to trade-off security for performance. \sagahyb~accommodates different level of security requirements for different type of agents by partitioning the user and agent registries on shards with different levels of security. Agents with higher level of privilege can run on byzantine-resilient \provider~shards, agents with medium-level of security requirements can ran on fault-tolerant \provider~shards augmented with auditing and monitoring, and lastly, the agents with small security risk can ran on \provider~shards that are fault-tolerant.

We evaluate all solutions and compare them against \saga. \sagabft~incurs significant overhead, achieving roughly 1\% of \saga's throughput for the most common class of requests. In contrast, monitoring and auditing achieve roughly 95\% and 85\% of \saga's throughput, respectively, on the same workload. Finally, we show that \sagahyb~amortizes the cost of byzantine resilience, incurring as little as $2\times$ the latency of \saga~while providing byzantine security for a chosen set of agents and users.
Our contributions are:
\begin{enumerate}
    \item We identify and demonstrate concrete attacks against SAGA's distributed \provider~architecture that compromise agent attributability, data confidentiality, and communication integrity.
    \item We design three types of solutions -- \sagabft, \sagamon~ and \sagaaud, and \sagahyb~-- that defend against these attacks while offering distinct security--performance trade-offs.
    \item We evaluate all solutions against SAGA, showing the conditions under which each solution is most effective.
\end{enumerate}

{\bf Paper roadmap}: Section \ref{sec:background} provides background on \saga. Section \ref{sec:threat_model} describes the attacker model for a malicious \provider~and the attacks we identify. Section \ref{sec:sagabft}, Section \ref{sec:m_and_a}, and Section \ref{sec:sagahyb} describe our solutions, \sagabft, \sagaaud~and \sagamon, and \sagahyb, respectively. We present the security analysis in Section \ref{sec:m_and_a_analysis} and our results in Section \ref{sec:results}.
We overview related work in Section \ref{sec:relwork} and conclude our paper in Section \ref{sec:concl}.

%% file: sections/2-background.tex
\section{Background}
\label{sec:background}

\subsection{SAGA Overview}
\label{sec:saga}

SAGA \cite{saga_ndss2026} is an architecture for secure governance of AI systems empowering users to have control over the lifecycle of their agents. A \texttt{user} is a human with a verifiable identity who owns and controls one or more \texttt{agents}, while an \texttt{agent} is an autonomous software designed to execute tasks, often relying on LLMs for decision-making. In SAGA, agents are cryptographically linked to the users who own them, and users create and control a contact policy that specifies what agents can interact with their agents and for how long. 

\begin{figure}[h]
    \centering
    \includegraphics[width=1.0\linewidth]{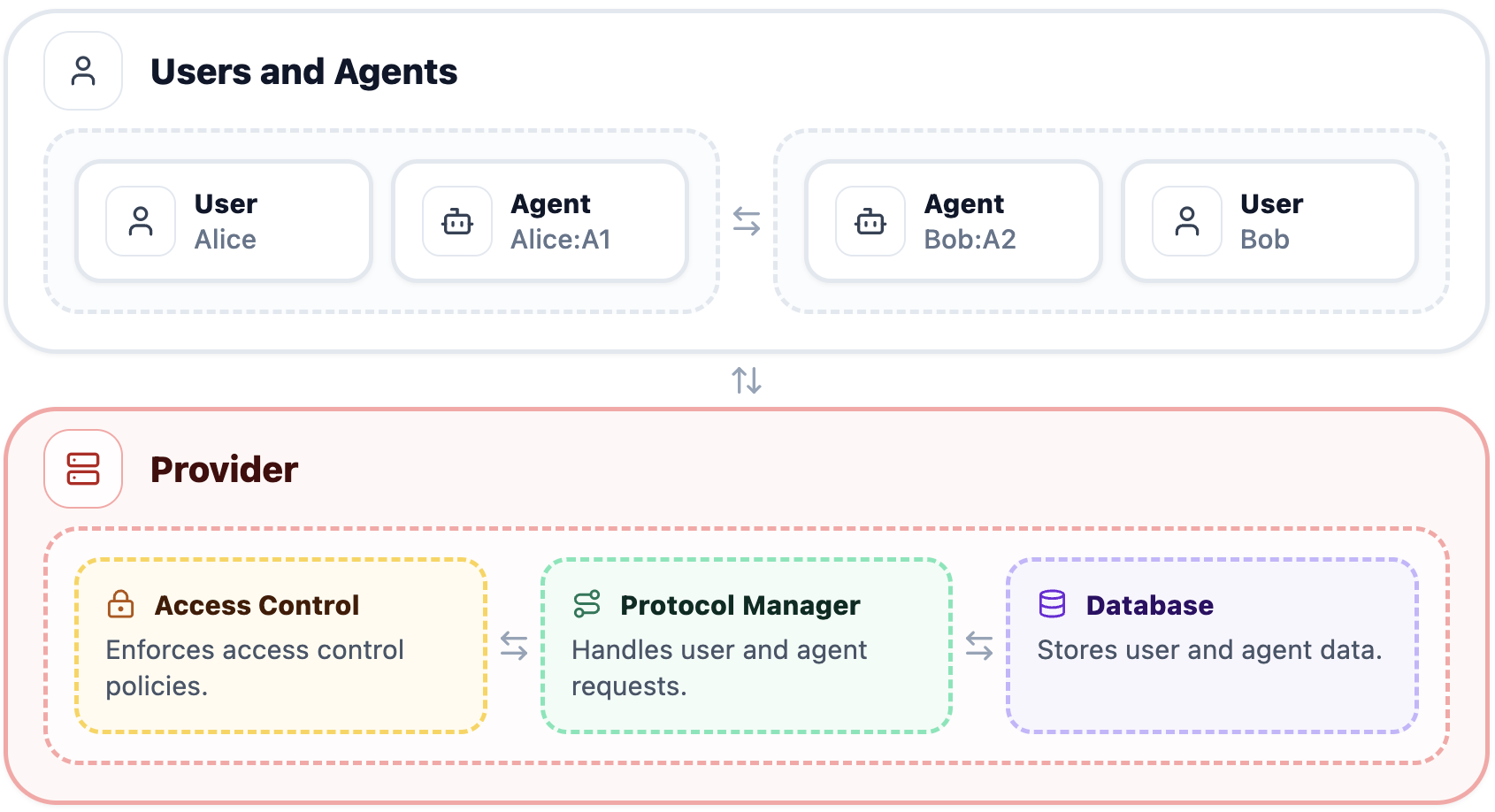}
    \caption{The SAGA architecture with two users, Alice and Bob, each owning agents A1 and A2, respectively. Users and Agents interact with the \provider~through the protocol manager which interfaces with the registries stored in a database and the access control engine. Agents also interact directly with each other as described in \ref{sec:saga}.}
    \label{fig:saga_basic}
\end{figure}

SAGA uses a logically centralized entity called \provider~ that plays two roles: (1) serves as a \textit{registry} for users and agents information, and (2) serves as an \textit{access control engine} enforcing the contact policy for each agent as specified by the user who owns it. (See Figure \ref{fig:saga_basic}).

Consider two agents, $A$ and $B$ that want to communicate. $A$ controls the access of $B$ to $A$ via an Access Control Token (ACT) generated by $A$ specifically for $B$ to use. However, before $B$ can talk with $A$, it needs to discover $A$ and it does this by contacting the \provider. The \provider~ will check if agent $B$ is allowed to contact agent $A$ and in the affirmative case will provide $B$ with information about how to contact $A$ as well as a one-time key (OTK) that will allow the two agents to establish a new secret key based on which the ACT will be derived. The ACT includes a validity timestamp and a maximum number of requests. When the ACT expires, $B$ must re-contact the \provider~ to obtain another OTK that it can use, by communicating with $A$, to compute a new ACT. 

\subsection{SAGA Protocols}
\label{sec:saga_protocols}
Users and agents interact with the \provider~through four protocols.

\textbf{User Registration}: This protocol registers a user with a real-world human identity so that agents can be linked to a human actor providing attributability.\footnote{In an OpenAI white paper Shavit et. al. define attributability as the ability to attribute the actions of an AI agent to a human user \cite{shavit2023practices}.} The \provider~verifies the user's identity using a service such as OpenID Connect \cite{sakimura2014openid}, then stores the user's data in the user registry.

\textbf{Agent Registration}: This protocol registers the information necessary for agents to establish communication. During registration, the user provides an \textit{agent card} containing the agent identifier (AID) and contact information -- both cryptographically bound to the user -- along with a set of OTKs and an \textit{Agent Contact Policy} (ACP) specifying the agent's permitted interactions. The ACP defines a set of declarative rules and an associated contact budgets, each specifying a pattern evaluated against the AID of a contacting agent. Upon registration, the \provider~initializes an \textit{access counter} to ensure the ACP's budget constraints are never violated.

\textbf{Agent Management}: This protocol allows users to modify their agents' ACPs, replenish OTKs, reset access counters, and revoke agents. The user specifies which agent and what artifacts are to be updated to the \provider. The \provider~then verifies the user owns the agent and modified the corresponding registry entry as instructed.

\textbf{Inter-Agent Communication}: This protocol facilitates secure conversation between agents. The contacting agent $A$ requests the information of the receiving agent $B$. The \provider~verifies access control by checking $A$  against $B$'s ACP and access counter, and ensures an OTK is available. If allowed, the \provider~forwards the signed contact information along with a single OTK. The agent can verify the signature on the contact information against the sender's public key to ensure it has not been tampered with. The OTK is used to derive an ACT, and the access counter is incremented.

\subsection{Fault-Tolerant and Scalable SAGA}
\saga~discusses several design ideas for fault-tolerance and scalability,
but it does not provide an implementation for them.

\begin{figure}[h!]
    \centering
    \includegraphics[width=1\linewidth]{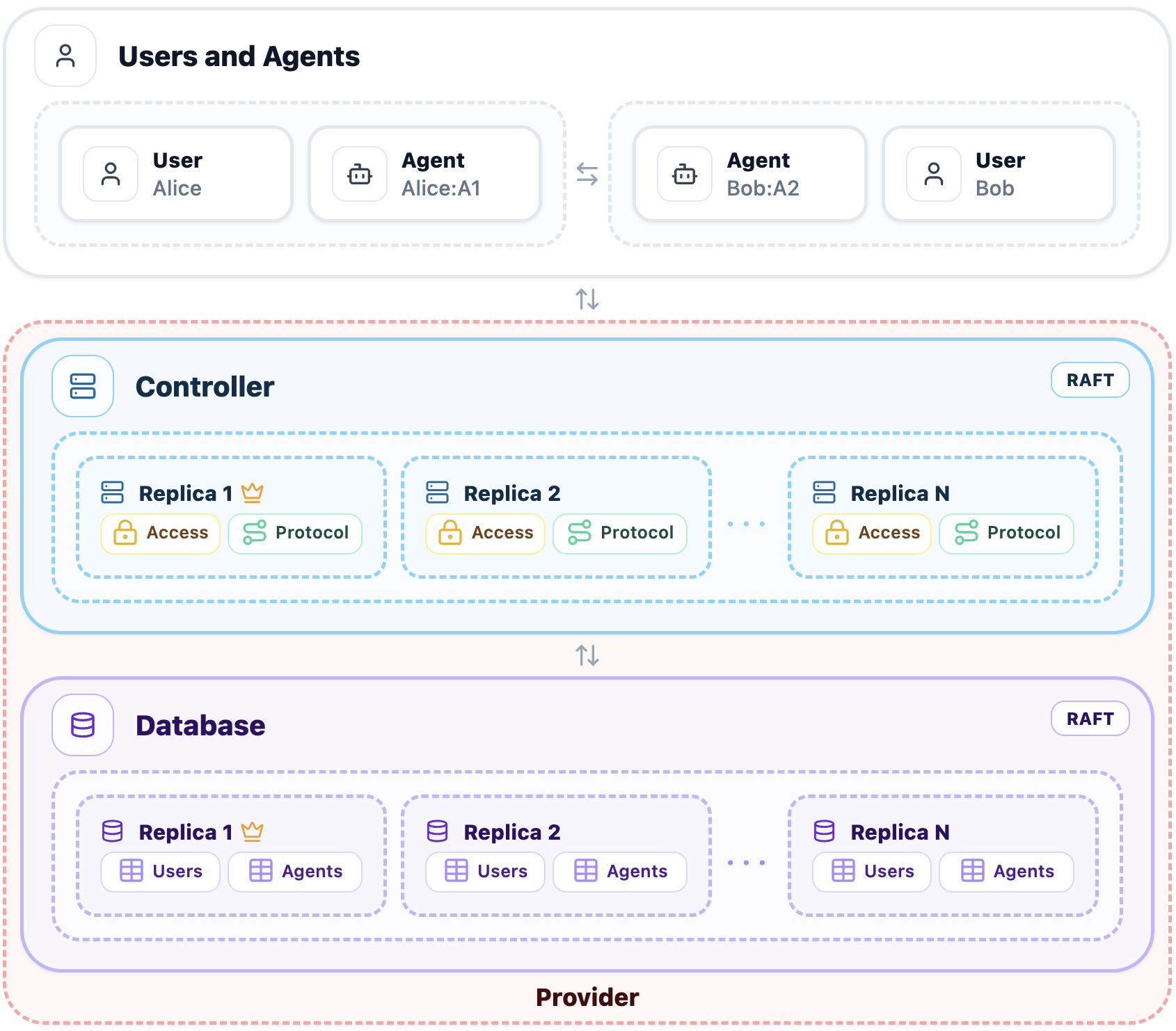}
    \caption{Fault-tolerant SAGA.}
    \label{fig:saga_ddb}
\end{figure}

\textbf{Fault-tolerance.} When deploying the \provider~in a real-world setting, it is essential to ensure that it is fault-tolerant, able to withstand system failures. To achieve this, \cite{saga_ndss2026} suggests distributing the \provider~using the Raft consensus algorithm. A \provider~replicated with the Raft consensus algorithm and $2f+1$ nodes can tolerate up to $f$ failures while preserving both availability and data integrity~\cite{ongaro2014search}. Given the availability of production-grade Raft-replicated databases and different system requirements of the components, a natural design uses two Raft groups. Thus, the \provider~is logically split into two pieces: the \textit{\controller}~that enforces access control and protocol management and the \textit{database} (see  Figure~\ref{fig:saga_ddb}).

\begin{figure}
    \centering
    \includegraphics[width=1\linewidth]{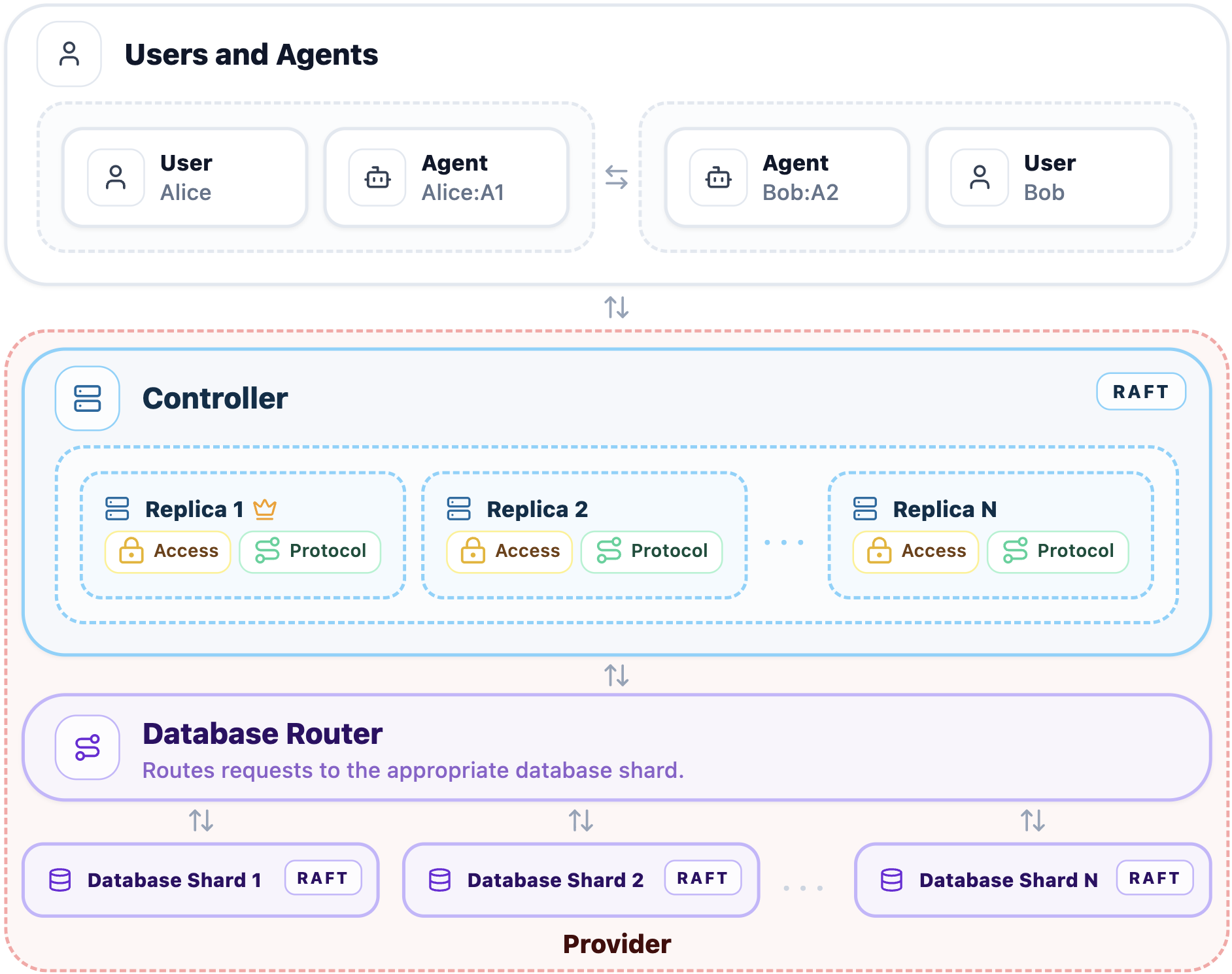}
    \caption{SAGA configured with a sharded database.}
    \label{fig:saga_dshard}
\end{figure}

\textbf{Scalability.} 
SAGA \cite{saga_ndss2026} discusses how to scale the database by using a common technique in databases called sharding \cite{chang2008bigtable,corbett2013spanner,baker2011megastore,korth2020database,solat2024sharding}. In sharding, data is partitioned uniformly across multiple independent database instances, known as \textit{shards}, each responsible for a subset of the overall data. A \textit{shard router} then routes queries to the correct shard. 
SAGA suggests a sharded database, connected to a single \controller~group, as depicted it Figure \ref{fig:saga_dshard}. 
Note that sharding works particularly well for SAGA because the partitioning can be done based on user or agent identifiers, respectively.

\subsection{SAGA Threat Model and Limitations}
\label{sec:saga_goals}

SAGA assumes a trusted provider model: the \provider~ is expected to faithfully execute all SAGA protocols and to preserve the confidentiality, integrity, and availability of both the user and agent registries. Under this assumption, the primary attack surface shifts to the agents themselves. SAGA's threat model considers adversarial agents that may deviate arbitrarily from prescribed protocols described in Section~\ref{sec:saga_protocols} -- including attempting to compromise other agents, exfiltrate data, escalate privileges, or replicate without authorization. The framework's defenses are designed to contain such misbehavior by leveraging the trusted \provider~as a centralized enforcement point for authentication, authorization, and policy compliance. 

\textbf{Limitations of a trusted Provider.}
While different configurations can improve SAGA's resilience to failures, SAGA relies on a trusted \provider~for correct and honest execution. Meeting fault-tolerance and scalability requirements necessitates extending the architecture to include external infrastructure such as replicated or sharded databases across a network of machines, which broadens the attack surface beyond the \provider~itself. 

%% file: sections/3-attacks.tex
\section{Attacks Against SAGA from a Malicious Provider}
\label{sec:threat_model}
In this section we consider the scenario when some components of the \provider~are compromised and controlled by an attacker. We describe the attacker model in \ref{sec:attacker_model}, then analyze several concrete attacks against the SAGA system.

\subsection{Attacker Model for a Malicious Provider}
\label{sec:attacker_model}

Different components of SAGA may be deployed in environments where they can be compromised \cite{ristenpart2009hey, jarkas2025container}. We assume the attacker's goal is to act in a byzantine manner, violating the correctness properties defined in Section~\ref{sec:saga_protocols}. We identify three distinct attack surfaces: the \textit{Access Control Engine} (ACE), the \textit{Protocol Manager} (PM), and the \textit{Database} (DB). The PM and ACE are logical separations of capabilities in the \controller. Recall that these are implemented as distributed services replicated with the Raft consensus protocol. We assume that in a system with $2f+1$ \controller~replicas and $2g+1$ database replicas, the attacker can compromise at most $f$ \controller~replicas and $g$ database replicas, including the leader. We focus on a single-shard configuration; however, the attacks naturally generalize to sharded deployments, with the caveat that an attacker cannot tamper with shards they do not participate in.

\begin{figure*}[t]
    \centering
    \includegraphics[width=1\linewidth]{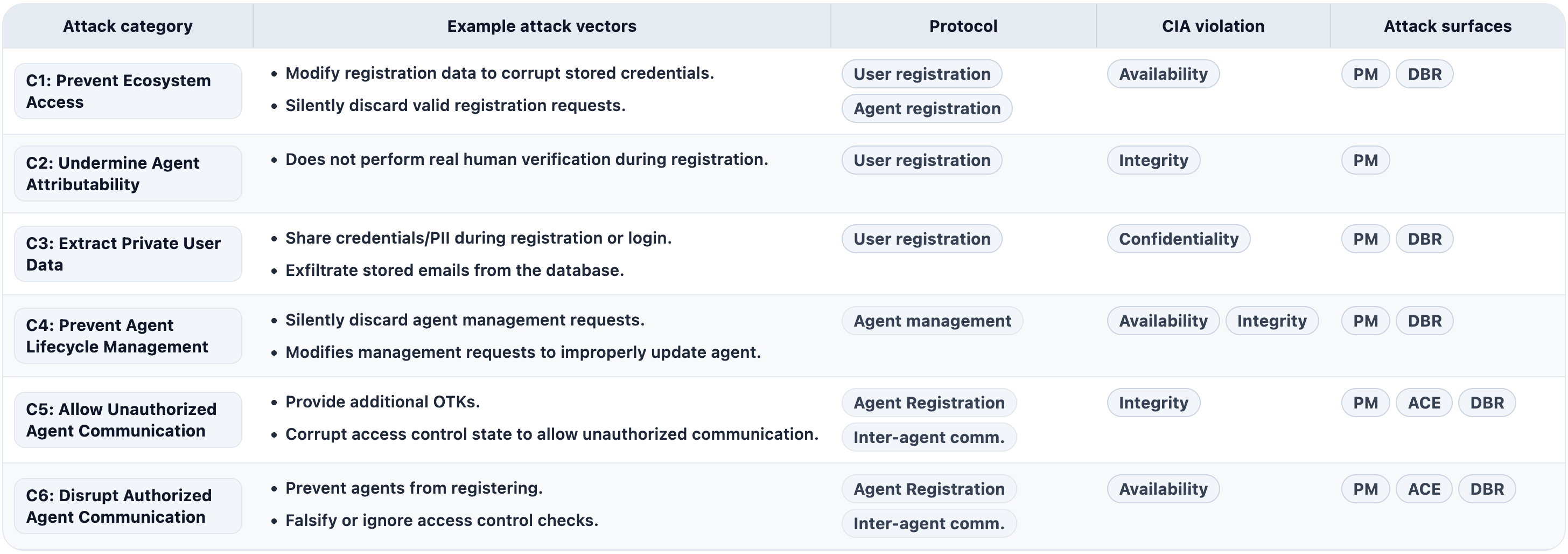}
    \caption{Summary of attack categories from a malicious \provider~against SAGA.}
    \label{fig:attacker_model}
\end{figure*}

A malicious PM is capable of deviating arbitrarily from the prescribed SAGA protocols -- including forging, dropping, or modifying messages. A malicious ACE can produce access control decisions that violate the specified policy. A malicious DB can violate the protocol by selectively omitting updates, tampering with them, or returning stale or fabricated responses.
We summarize the attacker model in Figure~\ref{fig:attacker_model} and describe each category in detail below.

\subsection{C1: Prevent Ecosystem Access} 

Preventing a valid user from registering, logging in, or registering agents denies them access to the ecosystem. A compromised PM or DB can achieve this through several methods -- either by modifying registration data to prevent the user from being correctly stored in the database, or by discarding requests entirely and returning a forged acknowledgment. This category of attacks constitute a violation of availability. While the impact is contained to the target user(s), the consequences can be permanent -- since SAGA does not permit duplicate users and each user is tied to a single human, even reinitializing with a trustworthy instance may not restore access if the corrupted registration occupies the user's unique identity slot.

\subsection{C2: Undermine Agent Attributability} \label{sec:attack:a2}

Attributability deters harm by allowing agent actions to be traced back to real, human identities \cite{shavit2023practices}. Circumventing attributability allows malicious agents to operate in the ecosystem without risk of consequence. This category violates the user registration protocol and can be executed by a compromised PM. By ignoring a failed response from an identity service, or skipping the request entirely, the PM can admit users to the system without binding them to a human identity. The impact of these attacks extends beyond a target user -- any agent in the ecosystem that can be contacted by the unattributed agent is at risk, as the agent can now act maliciously without consequences. These attacks violate integrity and can lead to confidentiality breaches. Recovery would be difficult as it would require all users re-verifying, or cross referencing all UIDs against the identity service's data.

\subsection{C3: Extract Private Data}
\label{attack:c3}
The \provider~ receives and stores sensitive user information, including email addresses and authentication credentials. A compromised PM can intercept this data during registration or authentication, and a compromised DB can extract personally identifiable information (PII) directly from the registry at any point after it has been stored. Compromised credentials can grant an attacker unauthorized access to the victim's agents or, if the user reuses passwords across services, to accounts on external platforms. Extracted email addresses expose users to identity-based attacks and targeted phishing. These attacks constitute a violation of confidentiality with a wide-ranging impact. Since the attacker obtains raw credentials and PII, the damage extends beyond the SAGA ecosystem and persists indefinitely regardless of whether the compromised component is subsequently detected and replaced.

\subsection{C4: Prevent Agent Lifecycle Management} Managing an agent once deployed is essential for safety. The ability to revoke an agent is a critical fail-safe for preventing both accidental and intentional harm \cite{shavit2023practices}, and serves as a last resort for blocking active attacks. Without this ability, rogue agents can cause unchecked harm. Conversely, refreshing OTKs and the access counter, as well as updating contact policies, allows an agent to persist safely in the ecosystem. Without maintaining these, an agent will become unreachable. Further, the ability to update contact policies, allows users to block a newly discovered malicious agent or enable communication with a new agents as required by an evolving workflow. This category of attacks violate the agent management protocol and can be realized by either a compromised PM or DB. Either component can silently block or tamper with modification requests from reaching the database while returning a false acknowledgment to the user, leaving them unaware that their changes were never applied. These attacks impacts are contained to the target user(s). If detected these attacks should be recoverable after reinitialization;  malicious contact during the vulnerability window cannot be undone. 

\subsection{C5: Allow Unauthorized Agent Communication}\label{attack:a5} 

Access control is essential to the SAGA ecosystem. Without enforceable access control, a malicious agent can discover or contact agents it is forbidden from reaching, enabling a variety of known attacks \cite{ferrag2025prompt, huang2024resilience, lee2024prompt, deng2025ai, zheng2025demonstrations}. Even if the target agent independently enforces its own access control, extracting the target's information and OTK during discovery enables the attacker to initiate future attacks under a different identity. These attacks can be executed by either a malicious ACE, PM, or DB and violates the inter-agent communication protocol -- the ACE by falsifying access control checks, the PM by ignoring correct ones, or the DB by returning falsified information or additional OTKs. The impact is limited to the targeted agent(s), but can result in significant and irreversible integrity and confidentiality violations. While access control can be restored after reinitializing the corrupted component, the damage to the target agent may already be done. Combining this category with category {\em Undermine Agent Attributability} (described in Section \ref{sec:attack:a2}), amplifies the harm, as the attacker can breach access control without consequences.

\subsection{C6: Disrupt Authorized Agent Communication} 

Conversely, inter-agent communication remains essential to multi-agent workflows. Preventing agents from reaching other agents they are authorized to contact may render them unable to execute their prescribed tasks. Improperly maintaining, initializing, or verifying the access counter, policy, or OTKs can lead to valid communication being rejected -- violating the inter-agent communication protocol. These attacks can be carried out by a compromised PM, ACE, or DB. The impact is in terms of availability and it contained to the target agent(s); however, blocked communication can lead to critical failures in important workflows. Reinitializing the corrupted component can reestablish communication flows, but errors that occurred during the disruption may be unrecoverable.

%% file: sections/4-bft.tex
\section{\sagabft: A Byzantine Fault Tolerant Architecture for SAGA}
\label{sec:sagabft}

We present \sagabft, a byzantine-resilient solution for SAGA.
For maximum security, both the \controller~and the database must be made byzantine fault-tolerant.

\subsection{Design}
\textbf{BFT Database.}
To ensure safe behavior in \saga's database we can replace the fault-tolerant database with a byzantine fault-tolerant database. Each database update is submitted to all replicas, which must reach agreement before the update is committed. The main difference is that BFT protocols require $3f+1$ nodes to tolerate up to $f$ byzantine faults \cite{pbft}, ensuring that the database operates correctly even in the presence of arbitrarily malicious replicas. Several blockchain database systems built on the Tendermint protocol \cite{buchman2018tendermint} already exist \cite{mcconaghy2016bigchaindb, el2019blockchaindb, ge2022hybrid}, providing well-tested BFT implementations that can serve as a foundation for \sagabft. Leveraging such established systems reduces both the engineering effort and the risk of implementation-level vulnerabilities. 

\textbf{BFT Controller.}
Making the \controller~Byzantine fault-tolerant is  more complex than the database. As with the database layer, the consensus algorithm would need to be upgraded from a fault-tolerant one to BFT. Unlike the database layer where established BFT systems can be adopted off the shelf, the \controller~is a custom component. Thus, its access control logic, protocol management, and state transitions would each need to be replicated and agreed upon across $3g+1$ instances. This could be accomplished by adapting a BFT replicated state machine such as PBFT~\cite{pbft}; however, generic BFT state machine replication incurs significant performance degradation compared to purpose-built protocols~\cite{singh2008bft}.

\subsection{Security and Performance Limitations}
\textbf{Security.}
The key advantage of a fully BFT configuration is that it is provably secure against all classes of integrity and availability attacks from compromised components of the \provider, as long as the BFT assumptions are met. This means no more than $f$ replicas are compromised in the case of the database, and no more than $g$ replicas are compromised in the case of the controller, for a system with $3f+1$ database replicas  and $3g+1$ controller replicas.   Confidentiality attacks \ref{attack:c3}, however, falls outside this guarantee: BFT protocols govern agreement and ordering, not what data nodes are permitted to observe or leak.

\textbf{Performance Limitations.}
The primary drawback of byzantine-resilient consensus is its significantly higher communication overhead compared to fault-tolerant consensus. Where Raft requires $O(n)$ messages per operation, BFT protocols often require $O(n^2)$ messages per operation due to the all-to-all communication rounds necessary to ensure agreement in the presence of malicious nodes. This quadratic overhead would directly impact the latency of every user and agent interaction with the \provider. Furthermore, correctly implementing a BFT protocol is notoriously difficult, and implementation bugs have historically led to the very vulnerabilities these protocols are designed to prevent~\cite{wong2024beyond, winter2023byzzfuzz}.

%% file: sections/5-monitor_and_audit.tex
\section{Mitigation by Monitoring and Auditing}
\label{sec:m_and_a}

In this section we describe lightweight solutions to mitigate the attacks we described in Section \ref{sec:threat_model} based on monitoring and auditing. We first give an overview of the approach and then describe our auditing and monitoring solutions in detail.

\subsection{Overview}
Detecting byzantine behavior is typically very challenging when the ground truth is not known, or there are no known invariants in the system.
In the case of the \provider~service, we observe that we can create opportunities for the ground-truth to be known.
Specifically, we use auditing to introduce in the system actions, for which we know the correct answer. The auditor will take an active role and insert canary actions to see if the audited party performs the action correctly. 

We also observe that there are invariants that we can derive, based on the application semantic, for example what changes should be reflected in the database, in response to a message sent over the network. Unlike the auditor, the monitor is a passive observer of the actions of the system. We derive system invariants and use monitoring to analyze database logs and network communication to check if the request and response observed match, as they are specified by the protocol. In our case, the target monitored are the different \provider~components (the \controller~and database) and can be done by the cloud operator where the \provider~is deployed. Note that the \provider~is implemented as a distributed service thus the monitoring implies monitoring all the nodes from the service.

Once a compromised component is detected, it can be reinitialized from a trusted image, restoring safe operation. Users can pause their agents while the system is repaired.

\subsection{Client-Side Auditing}
\label{sec:auditing}

\textbf{Goals.} In this approach, users proactively audit the \provider. This is done in an end-to-end manner: users directly or through their agents, periodically insert canary actions, such as registering a new agent, or updating information for an existing one,
and observe the answer they receive from one of the \provider~nodes. 

\textbf{Assumptions.} This approach operates with exactly the view of the system available to any user, requiring no additional information or privileged access. It allows users who may not trust their \provider~to verify the integrity of the governance system as it relates to their own agents. A single user, or a group of users, can run the auditing algorithm independently. We assume that a malicious \provider~cannot
distinguish the auditor from an ordinary user, and therefore
treats the auditor’s actions in the same manner as those of any
other participant.

\textbf{Algorithm.}
The algorithm operates on a cycle that can be manually initiated, run periodically, or triggered by a configurable set of rules or conditions. 
During each cycle the auditor user will insert a list of actions, and assert that the \provider~responses are consistent with correct protocol execution. The set of actions our algorithm uses are: \textcircled{1} registering a new user, \textcircled{2} registering and revoking an agent, \textcircled{3} registering two agents with policies that prevents mutual communication, and \textcircled{4} registering two agents with policies that allows communication. More details below:
\begin{itemize}
\item {\em Register a new user}: The auditor attempts to register a new user with invalid credentials. The user registration protocol requires the \provider~to reject such requests; if the registration succeeds, the algorithm concludes that the \provider~has been compromised.
\item {\em Register and revoke an agent}: The auditor probes the agent management endpoint by registering an agent they own and subsequently revoking it. The auditor then attempts to access the revoked agent; if the request succeeds, the revocation was not correctly executed, indicating a compromised \provider. 
\item {\em Prohibited communication}: The auditor registers two agents whose contact policies prohibit mutual communication; if the \provider~permits the interaction. it has been compromised. 
\item {\em Allowed communication}: The auditor registers two agents whose policies explicitly allow communication and verify that the \provider~correctly facilitates the exchange. A failure indicates that the \provider~is not faithfully enforcing the contact policy. 
\end{itemize}

For more effective auditing, users can randomize they auditing cycles and coordinate with other agents. This cycle assumes that attacks occur at the granularity of full protocols. More sophisticated attacks can be detected by augmenting the cycle with finer-grained checks that pinpoint which stage of the protocol is failing. See  Figure~\ref{alg:usa} in Appendix~\ref{app:ma_algos}.

\textbf{Discussion.}
The main limitation of this technique is that if the \provider~is able to distinguish between an auditor and a regular user, then the \provider~can behave correctly when audited, and malicious otherwise. 
To ensure that the \provider~cannot distinguish auditor clients from regular clients, auditors must behave exactly like normal users in every observable way. They should use the same client software and public APIs, avoid any special credentials or identifiable account traits, and match typical network characteristics such as IP type, TLS configuration, and request headers. Their activity should mimic real user behavior, including realistic timing, usage patterns, and variability, while being blended into a larger pool of similar clients to avoid standing out. Audits should be randomized in timing and actions.

 Another limitation is that, since auditing operates on the client side, in some cases there is no way to distinguish between a malicious controller and a malicious database. Thus, upon detecting an error further investigation is required. 
 
 Finally, similar to \sagabft~our defense focuses on correct execution of the protocol and does not address how an attacker handles private data (\ref{attack:c3}).

\subsection{Provider-Side Monitoring}

\begin{figure}[h!]
    \centering
    \includegraphics[width=1\linewidth]{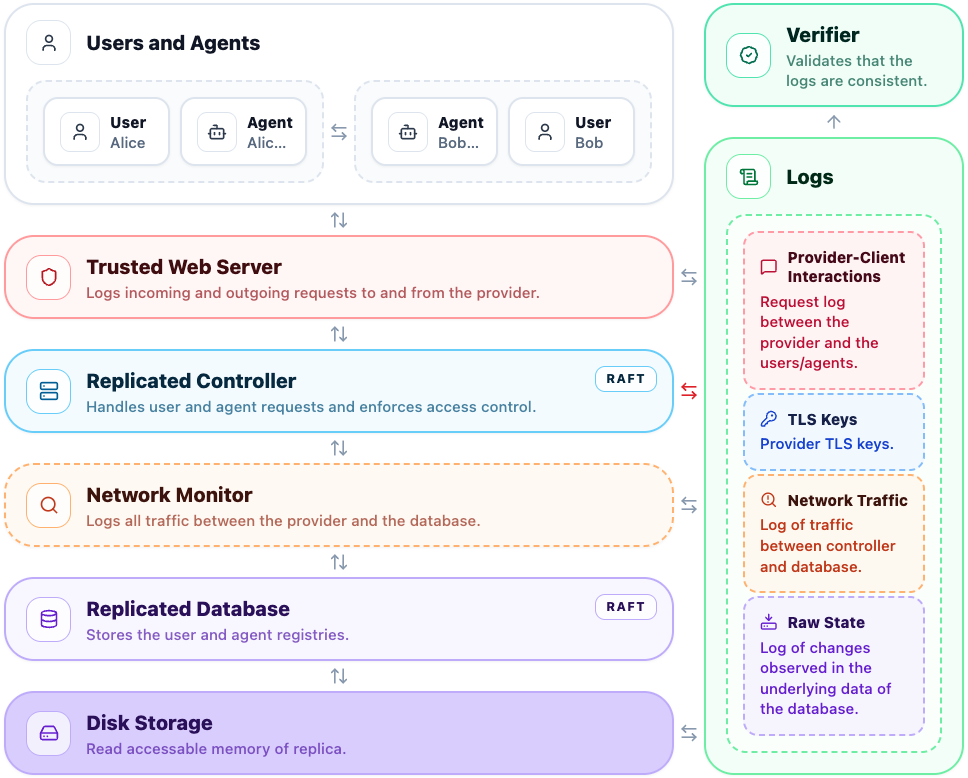}
    \caption{The monitoring architecture for \saga. Trusted components emit logs to a shared verifier, which correlates them to ensure consistency across the protocol. \controller~is not trusted; it is only responsible for dumping its TLS session keys, any withholding or forgery of which is detectable as undecodable traffic. The network monitor is optional; it attributes detected malicious behavior to a specific component.}
    \label{fig:saga_monitoring}
\end{figure}

\textbf{Goals.}
In this approach, the goal is to detect inconsistencies in the information transmitted and persisted across the various stages of the \saga~protocols. The monitoring service consists of log collection and a verifier that analyzes the logs, see Figure \ref{fig:saga_monitoring}.

\textbf{Assumptions.} 
We assume a cloud deployment where the \provider~service is distributed across multiple nodes, the database runs on separate and better protected nodes, and several trusted services are available to the cloud operator.
The monitoring algorithm is run by the cloud operator who is proactively monitoring the \provider~deployment. The monitor does not trust the \provider~code but it trusts two key sources of information: (i) a complete record of the requests and responses exchanged between the \provider~and its clients (users and agents), and (ii) a trusted, ground-truth view of the database state. With this information, the monitor can ensure integrity between the \provider~and supporting database. To verify integrity within the \controller~and database, a quorum of the replicas' logs for each component must agree. For example, assuming a cluster of $2f+1$ database replicas tolerating at most $f$ faults, a quorum consists of $f+1$ replicas.

\textbf{Monitoring and logs collection.}
One approach is to capture traffic to and from the \provider~through a trusted web server, CGI, or sidecar proxy that records incoming requests and outgoing responses before they reach the untrusted \provider~logic. Although client--provider traffic is protected by TLS, in most deployments TLS termination occurs at or before these layers, allowing the verifier to observe the messages unencrypted. An alternative approach is to monitor network traffic directly: the \provider~dumps its TLS session keys to the verifier, which then decrypts the traffic. Under this model, any attempt by a compromised \provider~to withhold or falsify its session keys would manifest as undecodable traffic and be treated as a system compromise. Traffic is stored in the logs as a tuple of requests and responses.

Obtaining the view of the database requires only read access to the underlying replica machines. While databases are typically accessed through a query language, their state is persisted in a raw on-disk format that can be read directly, bypassing any malicious logic in the database query engine. 

We can additionally monitor the network traffic exchanged between the \controller~and the database replicas. This enables the monitor to observe precisely which requests are issued by the \controller, making it possible to detect cases in which the \controller~fails to forward a client request to the database, issues spurious requests, or modifies requests in transit. Critically, this view allows us to attribute a detected inconsistency to either the \controller~or the database. We provide a full diagram of the setup in \Figure{fig:saga_monitoring}.

\textbf{Algorithm.}
To enable correlation across logs, the verifier attaches a unique action identifier to each transaction, allowing database operations to be matched to their originating client requests. The identifier is generated automatically when a client issues a request and requires no changes to client behavior. The \controller~records requests in first-in-first-out order, and the database likewise commits changes in order. The verifier can reconcile the two streams in a single ordered walk, advancing a pointer into each log in tandem. Periodically, it ingests completed request--response tuples from the \controller~and committed changes from the database, joined on the action identifier. Throughout, the verifier maintains an active model of the expected database state $\mathcal{D}$, advanced incrementally as each change is processed, so that suppression, injection, and tampering attacks are detected in a single traversal and access-control decisions can be audited against the reconstructed state at the moment each action was processed.

For this traversal, we introduce a window $W$ specifying how long the verifier waits before declaring an action unreconciled. $W$ represents the brief interval during which an action may legitimately be in flight between the \controller, the database, and the verifier. This same allowance, however, means that detection occurs on a short delay. In practice $W$ is bounded by network latency and is therefore small, so any sustained adversarial activity is detected quickly, before meaningful harm can be done. The algorithm is presented in detail in Figure~\ref{alg:log_verification} in Appendix~\ref{app:ma_algos}. When actively run on a fault-tolerant \provider, this configuration is referred to as \sagamon.

\textbf{Discussion.} One concern is the trustworthiness of the logs our framework relies on. If container-level logs cannot be trusted, TLS dumping with network-layer monitoring provides logs of \provider~behavior that a compromised component cannot tamper with.
Furthermore, even if container memory cannot be trusted, an initial database snapshot can be obtained from a quorum of replicas, and subsequent state can be reconstructed from the network logs as requests are processed.

%% file: sections/6-analysis.tex
\section{Security Analysis of Monitoring and Auditing}
\label{sec:m_and_a_analysis}
In this section, we formalize the security guarantees provided by our monitoring and auditing techniques. We characterize each mechanism in terms of the adversary it constrains, the assumptions it relies on, and the quantitative bounds it offers on detection. For auditing, we derive the expected time to detect a compromised \provider, the expected number of detections over a given time horizon, and the trade-offs that arise when the attacker and defender each tune their strategies. For monitoring, we formalize the invariants that bound the vulnerability window and analyze the trade-offs required to maintain them in practice. Together, these analyses delineate the configurations in which each mechanism is effective.

\subsection{Security Analysis of Auditing}
\label{sec:audit_analysis}
We assume a threat model in which the attacker cannot distinguish the auditor from an ordinary user, and therefore treats the auditor's actions in the same manner as those of any other participant. Under this assumption, the auditor's checks constitute a representative sample of the provider's behavior.

\begin{table}[h]
\centering
\small
\begin{tabular}{@{}cp{0.65\linewidth}@{}}
\toprule
\textbf{Variable} & \textbf{Definition} \\
\midrule
$m$      & Number of checks per audit cycle. \\
$\delta$ & Time between cycles. \\
$\alpha$ & Probability of attack, $\Pr(\text{attack})$. \\
$q$      & Detection probability given an attack, $\Pr(\text{detect} \mid \text{attack})$. \\
$D_T$      & Number of detections at time $T$. \\
\bottomrule
\end{tabular}
\caption{Auditing Variable Definitions}
\label{tab:audit-definitions}
\end{table}

We define our notation in Table~\ref{tab:audit-definitions}. We derive the expected time to detection. The probability that a single check both coincides with an ongoing attack and successfully detects it is
\[
\Pr(\text{attack}) \cdot \Pr(\text{detect} \mid \text{attack}) = \alpha q,
\]
assuming each check is independent, and the auditor performs $m$ checks per cycle, it detects an attack within a given cycle with probability
\[
s = 1 - (1 - \alpha q)^m.
\]
The cycle at which detection first occurs therefore follows a geometric distribution, $C \sim \mathrm{Geom}(s)$, with expectation $\mathbb{E}[C] = 1/s$. Each cycle has fixed duration $\delta$, so the expected time to detection $(\mathrm{TTD})$ is
\[
\mathbb{E}[\mathrm{TTD}] = \frac{\delta}{1 - (1 - \alpha q)^m}.
\]

Although the system should typically be reinitialized immediately upon detecting an error, it is also useful to consider the expected number of detections as a function of time. The number of detections across $n$ checks follows a binomial distribution,
\[
D_n \sim \mathrm{Binomial}(n, \alpha q),
\]
with expectation $\mathbb{E}[D_n] = n \alpha q$. In time $T$ the auditor performs between $\lfloor mT/\delta \rfloor$ and $\lceil mT/\delta \rceil$ checks. As we make no assumption about how checks are distributed within a cycle, we take the conservative approximation $n \approx \lfloor mT/\delta \rfloor$, yielding the expected number of detections in time $T$ ($D_T)$:
\[
\mathbb{E}[D_T] \approx \left\lfloor \frac{mT}{\delta} \right\rfloor \alpha q.
\]

Finally, we consider the detection--evasion game between defender and attacker, examining how each tunes their parameters to achieve their goals. For a small error rate, $\epsilon$, we ask how much time must elapse before detection occurs. This quantity is central to both sides: the attacker seeks to remain undetected with probability $1 - \epsilon$, while the auditor seeks to detect with the same probability. Recall that the probability of detecting an attack within a single cycle is $s = 1 - (1 - \alpha q)^m$, so the probability of at least one detection within $c$ cycles is
\[
\Pr(D_c > 0) = 1 - (1 - \alpha q)^{mc}.
\]
We know that $c \approx T/\delta$, thus we can equivalently express this in terms of time:
\[
\Pr(D_T > 0) = 1 - (1 - \alpha q)^{\frac{mT}{\delta}}.
\]
Separating the tunable parameters of each side, the defender most naturally tunes the cycle duration $\delta$, while the attacker tunes the attack probability $\alpha$. Solving $1 - \Pr(D_T) \geq 1 - \epsilon$ for the attacker and $\Pr(D_T) \geq 1 - \epsilon$ for the defender yields:
\begin{align*}
\alpha \leq \frac{1 - (1 -\epsilon)^{\frac{\delta}{(mT)}}}{q},
\qquad
\delta \leq \frac{mT \log(1 - \alpha q)}{\log \epsilon}. 
\end{align*}

\begin{figure}[h]
    \centering
    \begin{subfigure}[b]{0.48\linewidth}
        \centering
        \includegraphics[width=\linewidth]{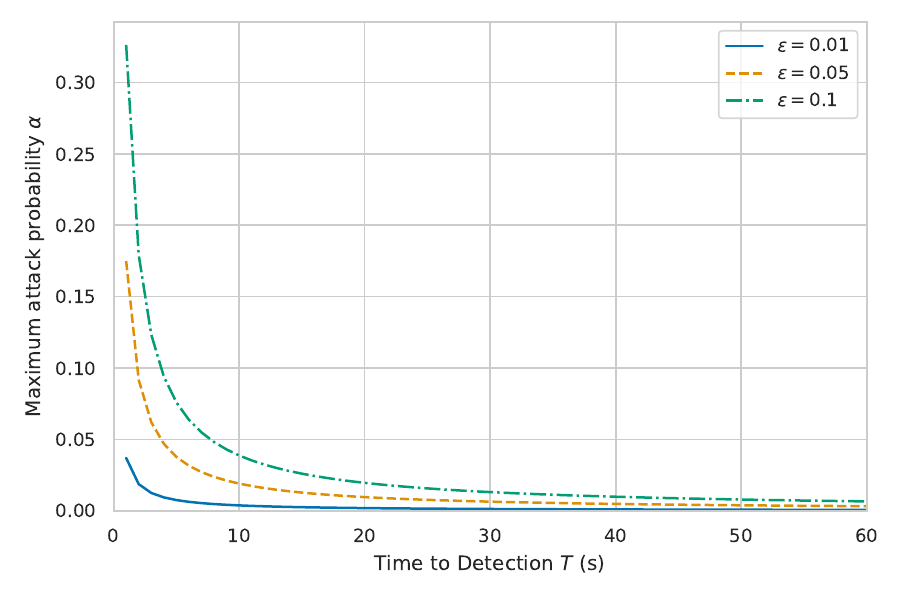}
        \caption{Attacker's bound on the attack probability $\alpha$ as a function of time $T$, for varying detection-confidence levels $\epsilon$. Each curve traces the maximum $\alpha$ at which the attacker remains undetected with probability at least $1 - \epsilon$ when $\delta = 15$.}
        \label{fig:attack}
    \end{subfigure}
    \hfill
    \begin{subfigure}[b]{0.48\linewidth}
        \centering
        \includegraphics[width=\linewidth]{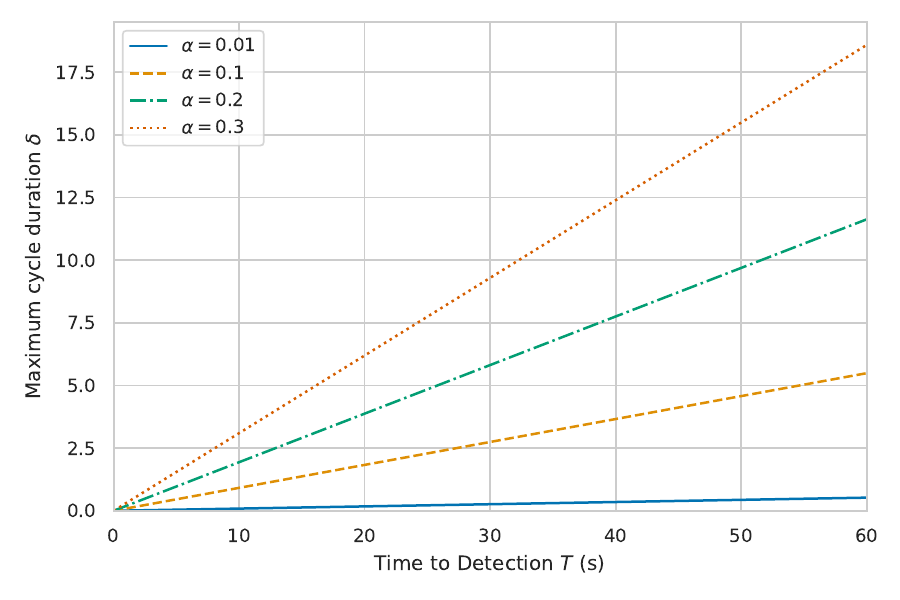}
        \caption{Defender's bound on the cycle duration $\delta$ as a function of the time to detection $T$, for varying attacker per-check attack probabilities $\alpha$. Each curve traces the maximum $\delta$ for which the auditor detects an ongoing attack with $\epsilon = 0.01$.}
        \label{fig:defense}
    \end{subfigure}
    \caption{Bounds for tuning parameters for the attacker and defender.}
    \label{fig:attack_and_defense}
\end{figure}

For our auditing cycle as described in Section~\ref{sec:auditing}, we set $m = 4$ checks per cycle and detection probability $q = 1$. Assuming the attacker can learn the defender's parameters, they tune their attack rate $\alpha$ based on an acceptable confidence level $\epsilon$ and attack window $T$ (Figure~\ref{fig:attack}). For this experiment we set $\delta = 15$. From the defender's perspective, we fix $\epsilon$ and show admissible cycle durations $\delta$ under varying attacker strategies and tolerated vulnerability windows (Figure~\ref{fig:defense}).

Under our configuration, an attacker must use $\alpha$ close to 0 to remain undetected for only 60 seconds, even when tolerating a high, 10\%, detection risk. Conversely, the defender needs frequent auditing to achieve 99\% detection against small $\alpha$, while larger $\alpha$ permits longer audit cycles.

\subsection{Security Analysis of Monitoring}
\label{sec:mon_analysis}
In this section we formalize the security guarantees provided by the \provider-side monitor. We assume a system with $2f + 1$ \controller~nodes and $2g + 1$ database nodes. An adversary $\mathcal{A}$ can control at most $f$ \controller~replicas and $g$ database replicas, and has no control over the verifier. $\mathcal{A}$ possesses all attack capabilities described in Section~\ref{sec:attacker_model}. We assume the verifier observes both logs in order through tamper-evident channels and that the action-identifier scheme is collision-resistant. We define \textit{actions} as the set of requests received by the \controller~and \textit{changes} as those sent to the database. The full set of variables used in our monitoring framework is defined in Table~\ref{tab:mon-definitions}.

\begin{table}[h]
\centering
\small
\begin{tabular}{@{}cp{0.65\linewidth}@{}}
\toprule
\textbf{Variable} & \textbf{Definition} \\
\midrule
$\mathcal{D}_i$ & Database state after action $a_i$. \\
$\mathcal{R}^*$ & Expected response given the database state. \\
$A = [a_1, \dots, a_n]$ & Sequence of all actions. \\
$C = [c_1, \dots, c_n]$ & Sequence of all changes. \\
$\Delta^* = \mathcal{D}_i - \mathcal{D}_{i-1}$ & Expected change resulting from action $a_i$. \\
$W$ & Window of potential matches. \\
\bottomrule
\end{tabular}
\caption{Monitoring variable definitions.}
\label{tab:mon-definitions}
\end{table}

For monitoring to remain coherent under replication, each cluster must expose a single canonical log to the verifier. Since both the \controller~and the database are replicated using Raft, this follows directly from the protocol's guarantees: Raft ensures that any committed entry is durably replicated to a quorum~\cite{ongaro2014search}. The canonical \controller~log ($A$) and database log ($C$) presented to the verifier are therefore the sequences of entries committed by their respective quorums. 

Let $\mathit{id}(a_i)$, $\mathit{req}(a_i)$, and $\mathit{res}(a_i)$ be the identifier, request, and response of action $a_i$, respectively. For correct monitoring of integrity errors, we must ensure that for every action and change, the corresponding identifier is observed and that the change matches the one induced by the action. A response is \emph{valid}, denoted $\mathcal{R}^*$, if it is consistent with both the returned status and the current database state. This consistency is essential for access control: if the response indicates success, the database state must reflect that the operation was permitted, and conversely, any change to the database must correspond to a successfully authorized action. We define these invariants:
\begin{align*}
&\textcircled{1} \quad\forall i\;:\; &\mathit{id}(a_i) = \mathit{id}(c_i) \;\;\wedge\;\; c_i = \Delta^*(a_i). \\
&\textcircled{2}\quad\forall i\;:\; ~&\mathit{res}(a_i) = \mathcal{R}^*\bigl(\mathcal{D}_{i-1},\, \mathit{req}(a_i)\bigr).
\end{align*}

The key challenge in enforcing these invariants is alignment: when $\mathcal{A}$ may drop, modify, or inject actions, the verifier must decide which entries to compare in the first place. We define $\hat{A} = [\hat{a}_1, \dots, \hat{a}_n]$ as the sequence of actions recorded in the \controller~log, and $\hat{C} = [\hat{c}_1, \dots, \hat{c}_m]$ as the sequence of changes read from the replica's disk. Both sequences preserve commit order, but neither is guaranteed to be complete: a legitimately delayed action may not yet appear in $\hat{C}$ at the moment $\hat{A}$ is read. To accommodate this benign skew without admitting adversarial drops, we introduce a window $W$ within which a corresponding entry must appear. We relax invariant \textcircled{1} into its windowed form, splitting it into forward and reverse directions to capture both suppression and injection; invariant \textcircled{2} is remains the same. 
\begin{align*}
&\forall\, \hat{a}_i \in \hat{A},\ \exists!\, \hat{c}_j \in \hat{C}_{[W]} : \mathit{id}(\hat{a}_i) = \mathit{id}(\hat{c}_j) \wedge \hat{c}_j = \Delta^*(\hat{a}_i), \\
&\forall\, \hat{c}_i \in \hat{C},\ \exists!\, \hat{a}_j \in \hat{A}_{[W]} : \mathit{id}(\hat{c}_i) = \mathit{id}(\hat{a}_j), \\
&\forall\, \hat{a}_i \in \hat{A} : \mathit{res}(\hat{a}_i) = \mathcal{R}^*\bigl(\mathcal{D}_{i-1},\, \mathit{req}(\hat{a}_i)\bigr).
\end{align*}

This raises the natural question of how to tune $W$. The window must be large enough that legitimately delayed entries are not mistaken for adversarial drops, yet small enough to bound the interval during which an attack may persist undetected. In practice, $W$ can be configured against a target false-positive tolerance (FP). Let $\mu$ and $\sigma$ denote the mean and standard deviation of the delay between the \controller~and the database in delivering an entry to the verifier. Modeling this delay as log-normally distributed -- a standard choice for request latencies~\cite{paxson2002empirically} -- the false-positive rate induced by flagging actions whose corresponding logs have not yet been fully populated is:
\[
\mathrm{FP}(W) = 1 - F_{\mathrm{LN}}(W;\, \mu,\, \sigma),
\]
where $F_{\mathrm{LN}}$ is the log-normal CDF. A target $\mathrm{FP} \leq \epsilon$ is therefore achieved by setting:
\[
W = F_{\mathrm{LN}}^{-1}(1 - \epsilon;\, \mu,\, \sigma).
\]

With this in mind, we can consider realistic deployments. Suppose the \controller~is located in US-East, with the verifier co-located with the \controller. We can then tune $W$ for different locations of the database. We fit a log-normal distribution to recorded AWS round-trip-times for databases located in US-West, Europe, and Asia, the resulting fit is shown in Figure~\ref{fig:latency_lognormal} in Appendix~\ref{app:ma_algos}.

%% file: sections/7-hybrid.tex
\section{\sagahyb~Architecture}
\label{sec:sagahyb}
This section presents \sagahyb, a natural extension to \sagabft~and \sagamon~that combines their respective strengths for large-scale deployments. \sagahyb~achieves the strong guarantees of \sagabft~for specific shards of the registry, while preserving the low overhead of \sagamon~throughout the rest of the system.

\subsection{Sharded Provider}
\label{sec:routing}

To increase the overall scalability and availability of the system, we  deploy multiple \provider~shards, each consisting of a replicated \controller~backed by its own database and responsible for a disjoint subset of the agent and user space. As shown in Figure~\ref{fig:saga_hybrid}, a routing layer is introduced through which all user and agent traffic is directed. The router maps each request to the appropriate \controller~based on the target user or agent. We refer to this composite system as \sagahyb. This architecture yields scalability benefits as the \provider~is no longer a computational bottleneck, and the system can scale horizontally by adding new \provider~shards.

\subsection{Heterogeneous Configurations}

\begin{figure}[h!]
    \centering
    \includegraphics[width=1\linewidth]{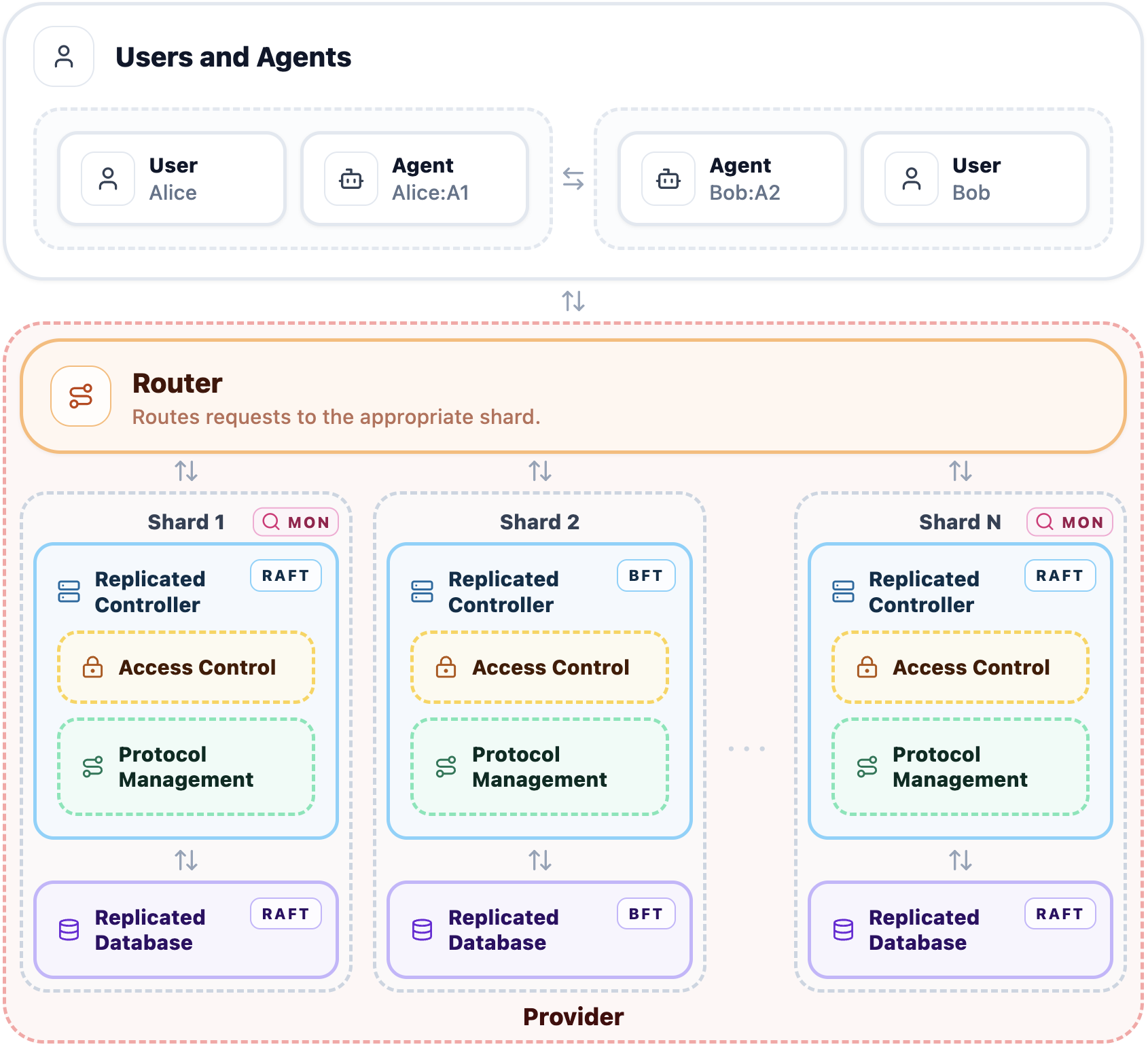}
    \caption{Hybrid configuration where some of the agent and user registry is run by a byzantine-resilient SAGA, while other are run by fault-tolerant SAGA some with monitoring enabled.}
    \label{fig:saga_hybrid}
\end{figure}

The monitoring and auditing techniques described in Section~\ref{sec:m_and_a} provide strong guarantees for detecting malicious behavior; however, detection is inherently delayed, and short vulnerability windows can still arise. In practice, different actors within the ecosystem may have varying requirements for the security--performance trade-off. Agents entrusted with highly sensitive data may warrant the stronger integrity guarantees of \sagabft, whereas latency-sensitive workloads may benefit from a database configuration with a large in-memory cache. The routing layer introduced in Section~\ref{sec:routing} (see Figure~\ref{fig:saga_hybrid}) naturally accommodates such heterogeneity: each \provider~shard can be independently configured with a distinct consensus protocol, database, caching policy, and monitoring cadence, while remaining accessible through a unified interface. This design enables the system to sustain high aggregate throughput while allocating stronger protections to the most sensitive subsets of state. This configuration is particularly well suited to large-scale, multi-tenant deployments, in which parties with divergent threat models and performance requirements must coexist within a shared infrastructure.

\textbf{Security.}
In \sagahyb, users and agents assigned to a BFT shard inherit the full safety guarantees of \sagabft, while all others fall under the protection of \sagamon. Furthermore, in \sagahyb~a single compromised \controller~or database can only affect the subset of data it is responsible for -- roughly $\frac{1}{n}$ of the total state, where $n$ is the number of \provider~shards -- limiting the radius of any individual compromise.

%% file: sections/8-experiments.tex
\section{Evaluation}
\label{sec:results}
In this section, we evaluate the Byzantine-resilient strategies for \saga~described in prior sections and quantify their effects on system performance and security. 

\subsection{Setup}
All experiments were conducted on a 16-core AMD Threadripper PRO 5955WX CPU paired with a Samsung MZ1L21T9HCLS-00A07 SSD. Agent workloads were driven by GPT-5.4, GPT-5.4-mini, and Qwen3-VL-30B-A3B-Instruct-FP8. Different nodes were deployed as containers and connected using a Docker network. We use \textit{Gunicorn} as our WSGI server, configured to handle multiple concurrent clients.

SAGA code does not implement fault-tolerance and scalability. We implemented the \saga~fault-tolerant and scalable architecture using the basecode of SAGA available at \cite{saga_git} and integrating it with RethinkDB~\cite{walsh2009rethinkdb}, a production-grade NoSQL database backed by a custom JSON store that natively supports replication and sharding.%

For \sagabft, we use BigchainDB~\cite{mcconaghy2016bigchaindb}, a BFT-backed key--value store built on top of the Tendermint (now CometBFT) consensus protocol~\cite{buchman2018tendermint}. Each registry action is modeled as a transaction comprising immutable data and mutable metadata. BigchainDB was shown to perform competitively among BFT-enabled databases~\cite{ge2022hybrid}. For \sagahyb, we additionally implement a custom routing layer that directs users and agents to the \provider~shard responsible for their state. Both RethinkDB and BigchainDB are configured to tolerate a single faulty replica -- requiring 3 and 4 replicas, respectively -- unless otherwise noted.

\subsection{Attacker Evaluation}

We evaluated SAGA's resilience against the adversarial behaviors outlined in our threat model (Section~\ref{sec:threat_model}). Our evaluation comprises 16 distinct attacks spanning all 6 attack categories. Attacks originate from each of the three system components -- protocol management (PM), access control engine (ACE), and database (DB) -- and  target all protocols, resulting in violations of confidentiality, integrity, and availability. We summarize the results of our attacks in Table~\ref{tab:launched_attacks} and provide a full description of all 16 attacks in Appendix~\ref{app:attacks}.

\begin{table}[h]
\centering
\small
\begin{tabular}{@{}lcccc@{}}
\toprule
 & \textbf{PM} & \textbf{ACE} & \textbf{DB} & \textbf{Total} \\
\midrule
C1: Prevent ecosystem access & 1 & 0 & 1 & 2 \\
C2: Undermine agent attributability & 1 & 0 & 0 & 1 \\
C3: Extract private user data & 1 & 0 & 1 & 2 \\
C4: Prevent agent management & 1 & 0 & 1 & 2 \\
C5: Allow unauthorized agent comm. & 1 & 1 & 2 & 4 \\
C6: Prevent authorized agent comm. & 2 & 1 & 2 & 5 \\
\midrule
 Confidentiality & 1 & 0 & 1 & 2 \\
 Integrity       & 4 & 1 & 3 & 8 \\
 Availability    & 2 & 1 & 3 & 6 \\
\midrule
 \textbf{Total} & 7 & 2 & 7 & 16 \\
\bottomrule
\end{tabular}
\caption{Distribution of executed attacks by originating component, broken down by attack category and security property violated. Some attacks violate multiple CIA properties; we report the most salient one in each case.}
\label{tab:launched_attacks}
\end{table}

\subsection{Monitoring and Auditing Evaluation}

To evaluate our monitoring and auditing frameworks, we implement a malicious \controller. Each request the \controller~processes is subjected to a probabilistic corruption model: with probability $\alpha$, the proxy behaves maliciously, modifying or suppressing the message in accordance with a configured attack strategy. The attacks are chosen are representative of each category mountable by a compromised \controller. The specific attack applied to each intercepted message is selected based on the protocol and direction.
We apply the attacks indiscriminately across all users and agents.

{\em Expected Number of Errors Detected by Auditing:} 
We configure our auditing protocol as described in Section~\ref{sec:auditing} with $m = 4$ checks per cycle and cycle duration $\delta = 15$. The attacker is configured with $\alpha = 0.3$, and each check has detection probability $q = 1$ when the corresponding component is attacked. Since we do not mount verification attacks, only 3 of the 4 checks are effective. As derived in Section~\ref{sec:audit_analysis}, with these parameters the expected number of detections by time $T$ is
\[
\mathbb{E}[D_T] \approx \left\lfloor \frac{3T}{15} \right\rfloor \cdot 0.3 = \left\lfloor \frac{T}{5} \right\rfloor \cdot 0.3.
\]

{\em Errors Detected with Monitoring:} 
We instantiate our monitoring hooks using a trusted CGI that intercepts client--\provider~traffic before it reaches the untrusted \provider~logic. For simplicity, we extract the database state by issuing queries from a whitelisted IP to the database infrastructure; a production implementation could instead read directly from the underlying on-disk data files to bypass the database query engine. 

Our monitoring protocol is configured with a window size $W = 100$\,ms and detects a wide range of malicious behaviors. We chose 100\,ms to effectively eliminate false positives in a US-East $\times$ US-West deployment (per Section~\ref{sec:mon_analysis}) and thus provide a faithful result. In some cases, the same attack can be flagged multiple times, or subsequent state mismatches can be detected in the aftermath of the initial violation. This redundancy is beneficial: since a single confirmed detection is sufficient to trigger system re-imaging, the goal is simply to catch the attack. However, for the analysis, we examine detection rate over time without triggering re-imaging; and accordingly, duplicate detections of the same underlying attack are filtered out.

\begin{figure}[h]
    \centering
    \includegraphics[width=1\linewidth]{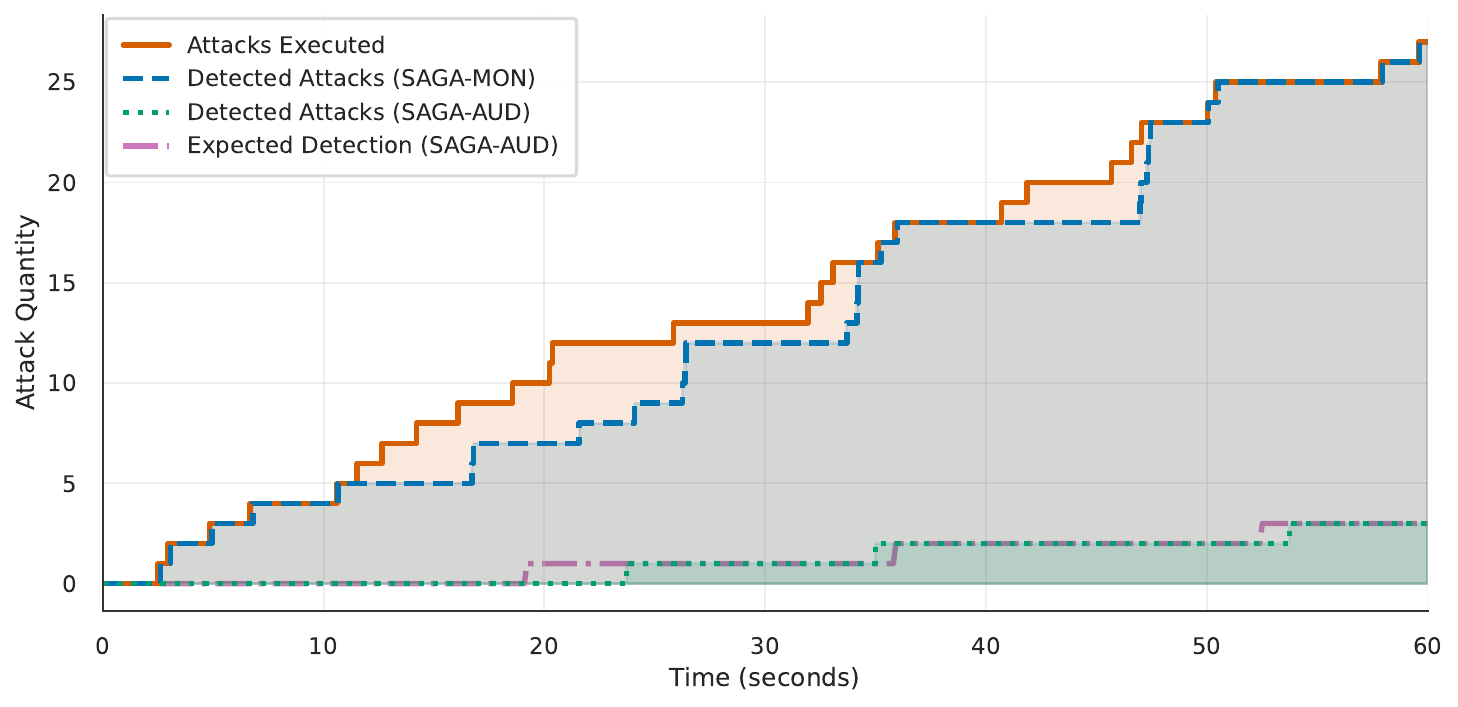}
    \caption{Detection rate of attacks introduced by the proxy over time. To generate this plot, we disable re-imaging on failure. We plot the expected number of detections for auditing as described in Section~\ref{sec:audit_analysis}, and use a window of $100$\,ms as motivated by Section~\ref{sec:mon_analysis}.}
    \label{fig:error_detection}
\end{figure}

\emph{Detection Analysis.}
As shown in Figure~\ref{fig:error_detection}, both auditing and monitoring rapidly detect compromises, surfacing multiple errors over the 60-second interval. The detections produced by auditing aligns with the analytically expected count, and monitoring detects every injected attack after a brief delay.

\subsection{Performance}
\label{sec:perf_experiments}
To quantify the trade-offs introduced by our solutions, we evaluate the end-to-end performance of \saga~-- in its vanilla configuration, with auditing enabled, and with monitoring enabled -- alongside \sagabft, across a range of workloads and deployment configurations. The vanilla configuration is fault-tolerant but does not perform monitoring or auditing. For these experiments, we configure auditing traffic to constitute 25\% of the total system load. We benchmark all configurations using only a replicated database. Since the database accounts for the majority of the cost, this provides a representative measure of the overhead.

\begin{figure}[h]
    \centering
    \includegraphics[width=0.7\linewidth]{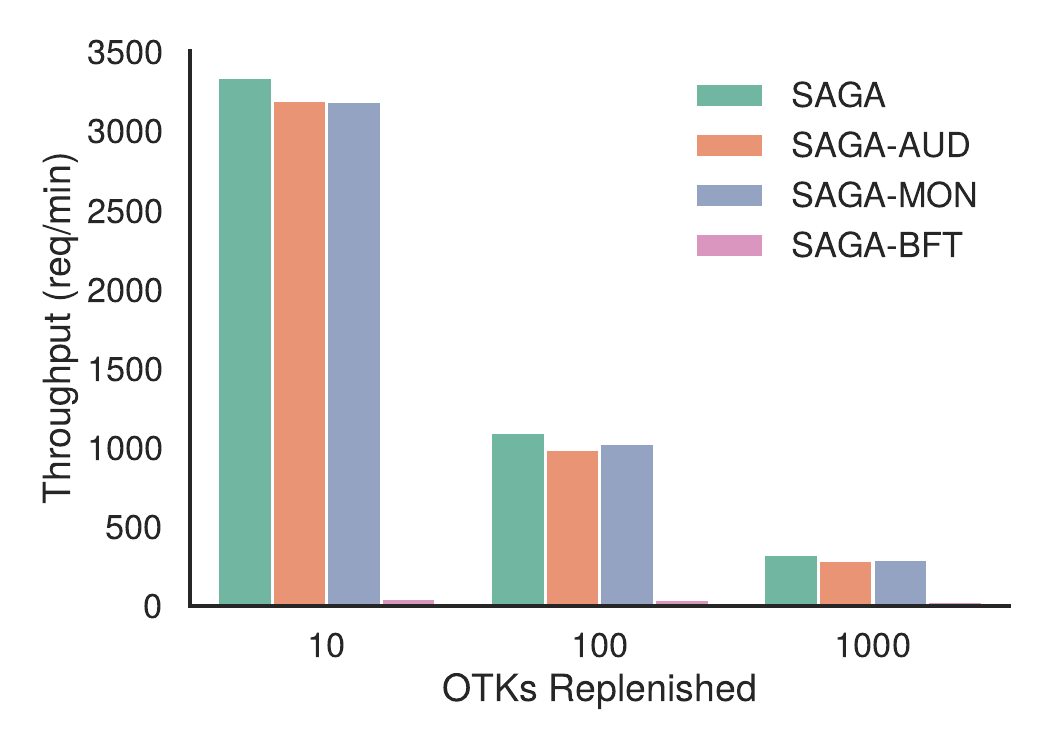}
    \caption{\provider~throughput handling OTK Refreshes.}
    \label{fig:otk_refresh}
\end{figure}

{\em OTK Refresh:}
We first evaluate the throughput of one-time key (OTK) refresh operations. OTK refreshes are a routine operation in \saga, ensuring that agents retain a sufficient pool of keys to service incoming contact requests. We measure throughput across refresh sizes of 10, 100, and 1000 tokens per operation, reflecting the fact that users can tune the refresh batch size according to the rate at which their agents consume keys. We report our results in Figure~\ref{fig:otk_refresh}.

As expected, lower token counts yield higher throughput across all configurations. \saga~achieves the highest throughput, while \sagaaud~and \sagamon~introduce only minimal overhead relative to the baseline. In contrast, \sagabft~incurs substantially higher latency, reflecting the inherent cost of Byzantine consensus. 

\begin{figure}[h]
    \centering
    \includegraphics[width=1\linewidth]{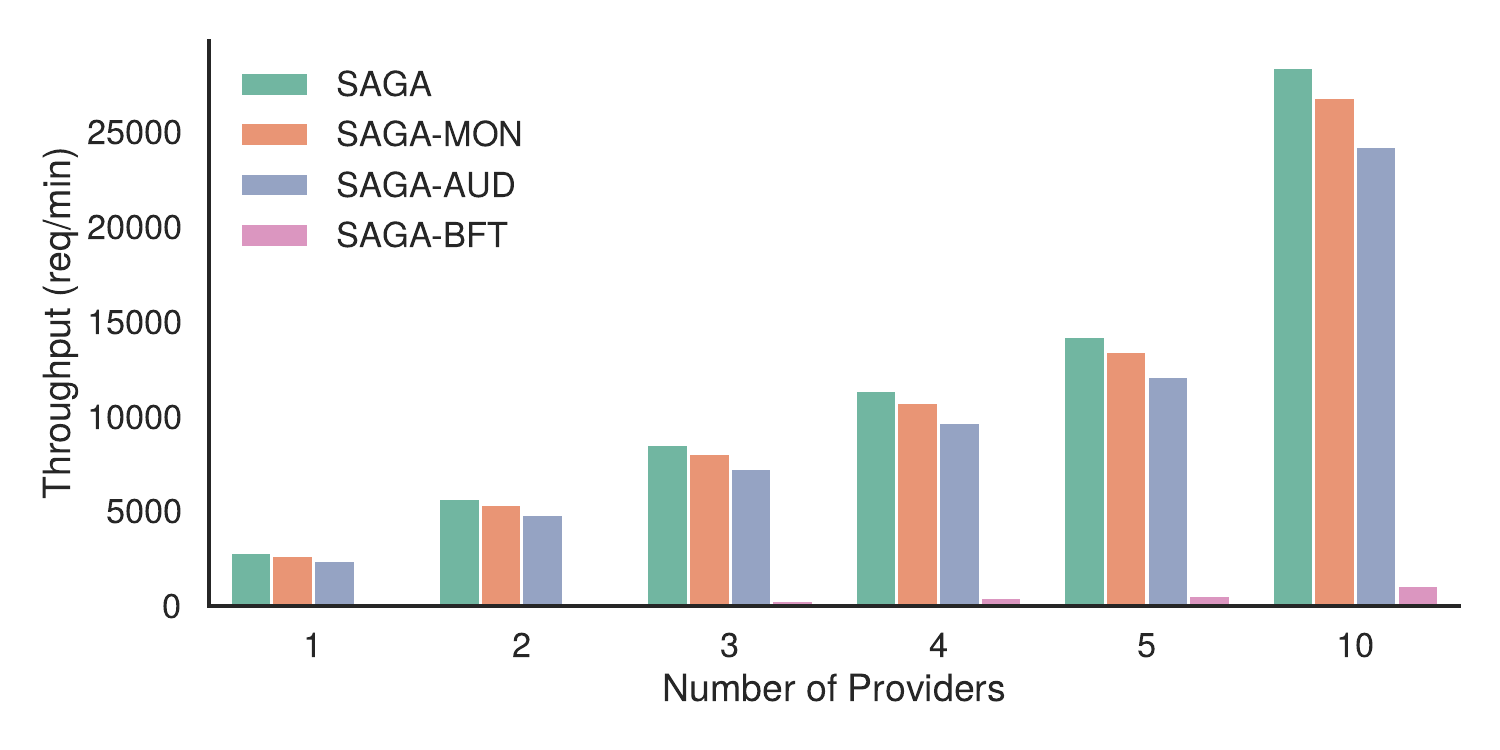}
    \caption{\provider~throughput handling inter-agent communication requests.}
    \label{fig:otk_access}
\end{figure}

{\em Inter-agent Communication:}
As described in Section~\ref{sec:saga}, initiating inter-agent communication requires an exchange with the \provider. In Figure~\ref{fig:otk_access}, we evaluate the number of such exchanges the system can sustain per minute. Consistent with the OTK refresh results, \saga~achieves the highest throughput, while \sagaaud~and \sagamon~introduce only minimal overhead. \sagabft, by contrast, incurs a substantial performance penalty. As before, the auditing load was configured at 25\%.

\begin{figure}[h]
    \centering
    \includegraphics[width=1.0\linewidth]{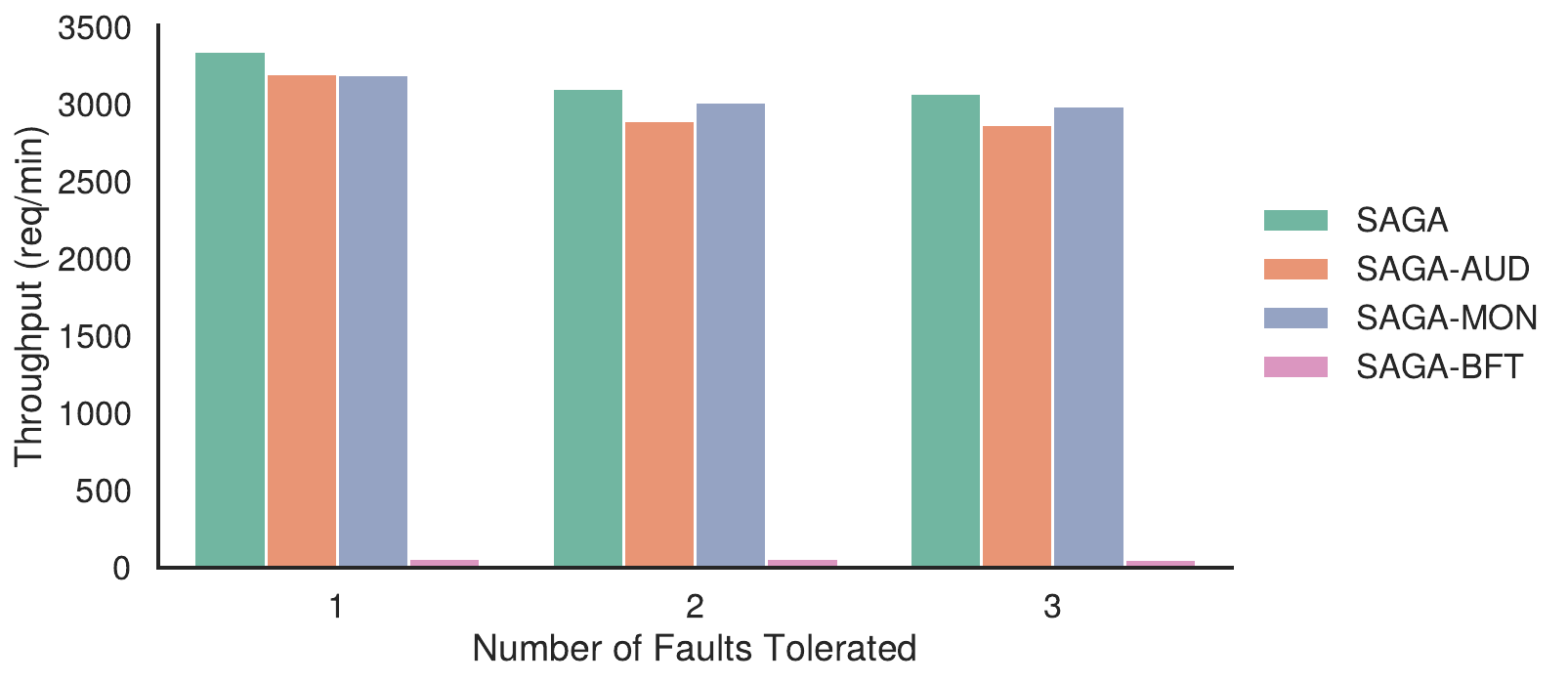}
    \caption{\provider~throughput handling different number of faults. \saga~requires $2f+1$ replicas and \sagabft~requires $3f+1$ where f is the number of faults.}
    \label{fig:number_of_faults}
\end{figure}

{\em Tolerating more Faults/Compromises:}
Thus far, our experiments have considered database configurations that tolerate a single faulty replica. We now evaluate throughput when the system is provisioned to tolerate 1, 2, and 3 simultaneous faults, requiring 3, 5, and 7 replicas for the Raft replicated and 4, 7, and 10 replicas for \sagabft, respectively. The results are presented in \Figure{fig:number_of_faults}. As the fault tolerance threshold increases, absolute throughput degrades slightly across all configurations; however, the key takeaways remain consistent with our previous experiments.

\begin{figure}[h]
    \centering
    \includegraphics[width=1\linewidth]{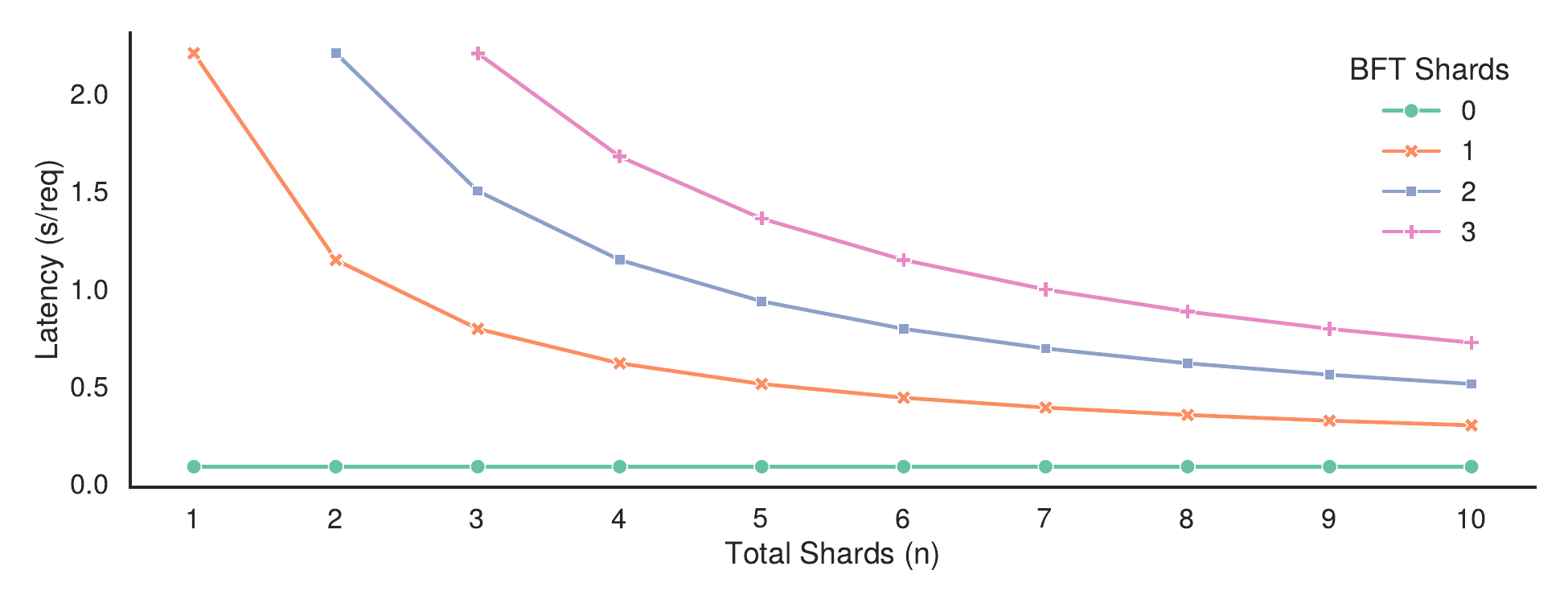}
    \caption{Latency of \sagahyb~with different number of \sagabft~shards. Non-BFT shards are using \sagamon.}
    \label{fig:hetero_shards}
\end{figure}

{\em \sagahyb~Configurations:}
Leveraging the \sagahyb~configuration, we can deploy hybrid systems in which some shards run \sagamon~while others run \sagabft. In Figure~\ref{fig:hetero_shards}, we report the average throughput of such heterogeneously configured \sagahyb~deployments. The reported latency is computed by adding the round-trip time of the router request to the average latency of each \provider~instance in the configuration; formally,
\[
\text{Latency} = \frac{\text{RTT}_{\text{router}} + (n - b) \cdot T_{\text{MON}} + b \cdot T_{\text{BFT}}}{n},
\]
where $n$ is the total number of shards, $b$ is the number of BFT-backed shards, and $T_{\text{MON}}$ and $T_{\text{BFT}}$ are the average inter-agent communication latencies for \sagamon~and \sagabft, respectively. As $n$ grows, the latency contribution of the BFT shards is amortized across the full set of providers, allowing the system to sustain high aggregate throughput even when a subset of shards runs under stronger consensus guarantees.

\subsection{End-to-end Evaluation}
\label{sec:endtoend}
To evaluate how our frameworks fare in a realistic multi-agent setting, we set up three tasks: (1) scheduling a meeting, (2) submitting an expense report for a trip, and (3) writing a collaborative blog post. We consider a highly distributed deployment with agents and \provider~components spread across US-East, US-West, Europe, and Asia. We measure the latency of each action, broken down into LLM thinking time (LLM), inter-agent network latency after the ACT is derived (Inter-Agent Network), and the latency overhead of each configuration relative to \saga. 
For results, see Table~\ref{tab:llm-latency-breakdown}.

\begin{table}[h]
\centering
\small
\begin{tabular}{@{}lccc@{}}
\toprule
& \textbf{Calendar} & \textbf{Email} & \textbf{Writing} \\
LLM Backend & \texttt{GPT-5.4-mini} & \texttt{GPT-5.4} & \texttt{Qwen-3} \\
\midrule
\multicolumn{4}{@{}l}{\textit{Standard Costs (s)}} \\
LLM        & 8.348 & 18.607 & 34.27 \\
Inter-Agent Network    & 0.563 & 0.676 & 0.901 \\
\midrule
\multicolumn{4}{@{}l}{\textit{Overhead (s)}} \\
\saga      & 0.323 & 0.323 & 0.323 \\
\sagamon   & 0.328 & 0.328 & 0.328 \\
\sagaaud   & 0.338 & 0.338 & 0.338 \\
\sagabft   & 2.452 & 2.452 & 2.452 \\
\sagahyb$_{1,10}$   & 0.632 & 0.632 & 0.632 \\

\bottomrule
\end{tabular}
\caption{Latency breakdown across LLM tasks, backends, and \provider~configurations. Each task is between two agents $A$ (initiator) and $B$ located in Europe and Asia respectively. The \controller~is located in US-East and the supporting database in US-West. All database replicas are assumed to be on the same LAN. \sagahyb$_{1,10}$ is configured with 1 BFT shard and 9 monitoring with the router located in US-East.}
\label{tab:llm-latency-breakdown}
\end{table}

As shown in Table~\ref{tab:llm-latency-breakdown}, the latencies of \sagamon~and \sagaaud~are comparable to \saga. \sagabft~incurs higher latency but remains reasonable for security-critical deployments when compared against LLM thinking time. \sagahyb~strikes a middle ground, providing \sagabft-level security where needed while keeping average latency on par with baseline inter-agent communication. Full transcripts of these experiments are provided in Appendix~\ref{app:transcripts}.

%% file: sections/9-related_work.tex
\section{Related Work}
\label{sec:relwork}

MCP enables agents to communicate through a standardized client-server architecture, in which an MCP server exposes capabilities that allow agents to interact with one another~\cite{anthropic2024mcp}.
A2A~\cite{google2025a2a} adopts a discovery-oriented approach, using \emph{agent cards} published at well-known endpoints to advertise an agent's capabilities and enable communication. Agora~\cite{marro2024agora} provides a negotiation-based communication layer that allows agents built on different models to interoperate. ANP~\cite{chang2025agent} relies on decentralized identity and authenticated messaging. These protocols remain susceptible to a range of threats~\cite{anbiaee2026security}.

{\em Attacks on Multi-Agent Systems:}
Ferrag et al.~\cite{ferrag2025prompt} provide an overview of the multitude of attacks that can disrupt AI agentic workflows. Huang et al.~\cite{huang2024resilience} study the resilience of MAS when a subset of agents behave faultily or adversarially, characterizing how such failures degrade overall system performance. Lee and Tiwari~\cite{lee2024prompt} investigate LLM-to-LLM prompt injection propagation among a set of LLM-based agents, like a virus. Kavathekar et al.~\cite{kavathekar2025tamas} introduce a benchmark for systematically measuring MAS vulnerabilities across a variety of attack vectors.

{\em Secure Protocols for Agent Communication:}
TrustAgent~\cite{hua2024trustagent} introduces a constitution-based defense in which agents are constrained by a set of declarative safety rules, but the approach offers no formal guarantees of enforcement.
Tsai and Bagdasarian~\cite{tsai2025contextual} propose a just-in-time security policy determination mechanism that dynamically derives access-control decisions from context.
Another line of work~\cite{louck2025improving, habler2025building, hou2026smcp} builds security principles onto existing communication protocols through trusted component registries, tamper-resistant logging, and authenticated endpoint discovery; however, similar to \saga, these approaches rely on a centrally trusted registry as a root of trust.

{\em Distributed Protocols:}
The AGNTCY Agent Directory Service~\cite{muscariello2025agntcy} implements a decentralized agent registry built on a distributed hash table (DHT), enabling peer-to-peer discovery without a central authority; however, the design does not account for the possibility of compromised nodes participating in the DHT. Huang et al.~\cite{huang2025fortifying} propose an architecture to defend against logic-level threats in agent interactions, but their approach relies on a trusted Agent Name Service to mediate discovery and authentication, reintroducing a centralized trust assumption that our work explicitly seeks to eliminate.

%% file: sections/10-conclusion.tex
\section{Conclusion}
\label{sec:concl}
In this paper, we establish and categorize the vulnerabilities that arise when an agentic governance service is untrusted or faulty, and demonstrate how these risks are amplified when such a service is deployed in a distributed setting. We first present a series of attacks that enable unsafe agent communication and can lead to catastrophic outcomes for both users and downstream systems. We then introduce \sagabft, \sagamon, and \sagaaud, three types of solutions for defending against malicious adversaries, and analyze the security--efficiency tradeoffs each entails. Building on these designs, we present \sagahyb, a unified solution that accommodates a wide range of deployment requirements. Taken together, this work demonstrates how to mitigate harm to users of these increasingly critical services even when individual components can be compromised.

%% file: sections/A-appendix.tex
\appendix%

\input{sections/A1-ma_algos}
\input{sections/A2-attacks}
\input{sections/A3-transcripts}

%% file: sections/A1-ma_algos.tex
\subsection{Additional Details for Auditing and Monitoring}
\label{app:ma_algos}
\begin{figure}[h]
   \centering
   \includegraphics[width=0.75\linewidth]{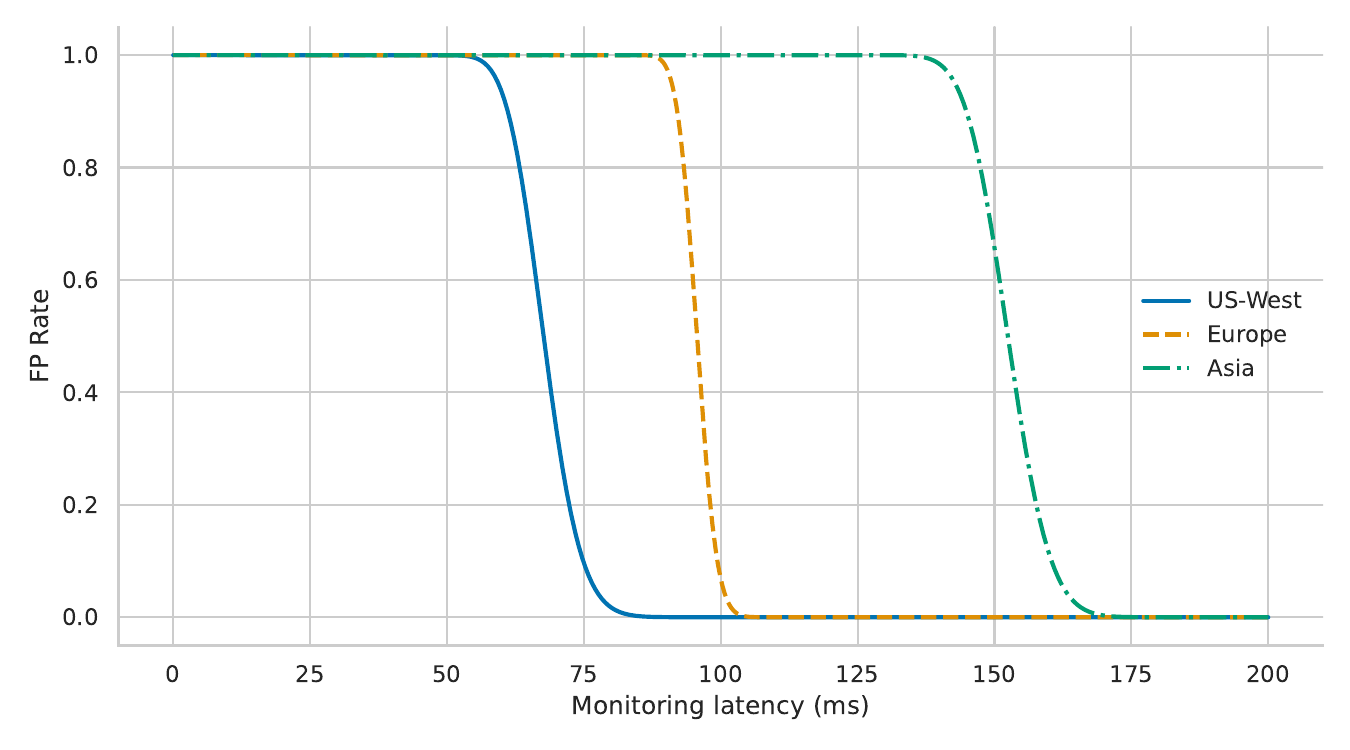}
   \caption{False-positive rate as a function of $W$ across different deployment locations of the database. The \controller~and verifier are both on US-East.}
   \label{fig:latency_lognormal}
\end{figure}

In this appendix, we first provide a detailed step-by-step description of the monitoring verification process. We then present the algorithms for auditing and monitoring described in Section~\ref{sec:m_and_a}, shown in Figures~\ref{alg:usa} and~\ref{alg:log_verification}, respectively. Lastly we present false-positive rates as a function of window size for various deployment locations in Figure~\ref{fig:latency_lognormal}. 

\textbf{Verification.}
\begin{itemize}
\item \textit{Step 1: Matching successful actions to database effects.}
For each action in logs$_{\mathit{ct}}$ that returned a success status, the verifier scans forward in logs$_{\mathit{db}}$ for a change carrying the same action identifier. Any database change encountered before the match is, by construction, unjustified: no client request produced it. Such changes are flagged as injected writes, establishing the invariant that \emph{every persisted state change corresponds to a valid, client-initiated action}. The verifier scans for a bounded interval $W$ before concluding that an unmatched request was not merely delayed but intentionally dropped.

\item \textit{Step 2: Verifying faithful execution of a match.}
When an action and a database change are matched, the verifier checks that the resulting row faithfully reflects the request. For user and agent registration, the stored row must exactly match the request the \controller~received. For agent management operations, the delta between the pre-image and post-image must correspond exactly to the requested mutation. For access operations, exactly one one-time key must be consumed and the action counter must be correctly decremented.

\item \textit{Step 3: Verifying access control decisions.}
For each inter-agent communication request, the verifier checks the access-control decision against $\mathcal{D}$, the reconstructed database state. For every allowed interaction, the verifier checks -- against $\mathcal{D}$ -- whether the target's contact policy, access counter, and one-time-key inventory truly permitted the request, and whether the response returned to the client is consistent with that state. For every denied interaction, the verifier checks that at least one of these conditions \emph{must} have failed. A violation in either direction constitutes a detected attack.

\item \textit{Step 4: Advancing the reconstructed state.}
After each change is processed, the verifier applies it to $\mathcal{D}$ so that subsequent access-control checks are evaluated against the current view of the system.

\item \textit{Step 5 (Optional): Attribution of failures.}
Upon detecting an inconsistency, the verifier can optionally examine the logs exchanged between the \controller~and the database to determine which component produced the falsified information. This enables precise attribution of the violation, allowing the system to re-image only the compromised component rather than the entire stack.
\end{itemize}

\begin{figure}[h]
    \centering
    \begin{algorithmic}
    \STATE \textbf{Algorithm:} \textsc{ClientSideAudit}(UID)
    \IF{UID not in registry}
        \STATE \textsc{RegisterProbeUser}(UID) \COMMENT{C1}
    \ENDIF
    \STATE \textit{// Run audit manually, periodically, or on trigger}
    \WHILE{\textbf{true}}
        \STATE \textsc{AuditUserVerification}( ) \COMMENT{C2}
        \STATE \textsc{AuditAgentManage}( ) \COMMENT{C4}
        \STATE \textsc{AuditInvalidAgentComm}(UID) \COMMENT{C5}
        \STATE \textsc{AuditValidAgentComm}(UID) \COMMENT{C6}
    \ENDWHILE
    \STATE \textbf{end function}
    \end{algorithmic}
    \caption{Algorithm for user side auditing.}
    \label{alg:usa}
\end{figure}

\begin{figure}[h]
    \centering
    \begin{algorithmic}
    \STATE \textbf{Algorithm:} \textsc{ProviderSideMonitoring}($W$)
    \STATE $\mathcal{D} \gets$ initial state from database
    \STATE $\textbf{A} \gets$ $\varnothing$; $\textbf{C} \gets$ $\varnothing$
    
    \LOOP
        \STATE append new actions from log$_{\mathit{ct}}$ to $\textbf{A}$
        \STATE append new changes from log$_{\mathit{db}}$ to $\textbf{C}$
        \FORALL{actions $a \in \textbf{A}$}
            \IF{$a$ succeeded}
                \STATE scan forward in $\textbf{C}$ for change $c$ matching $a$
                \IF{found}
                    \IF{$\textsc{AccessNotPermitted}(a, \mathcal{D})$}
                        \STATE \textcolor{red}{\textbf{flag}} access-control elevation
                    \ENDIF
                    \IF{$\textsc{IntegrityViolation}(a, c)$}
                        \STATE \textcolor{red}{\textbf{flag}} action tampered
                    \ENDIF
                    \STATE \textcolor{red}{\textbf{flag}} any skipped changes as \textit{injected}
                    \STATE $\textsc{UpdateState}(c, \mathcal{D})$
                    
                \ELSIF{scan exceeded $W$}
                    \STATE \textcolor{red}{\textbf{flag}} action suppressed
                \ELSE
                    \STATE defer $a$ to next cycle; \quad \textbf{break}
                \ENDIF
            \ELSIF{$a$ was denied}
                \IF{$\textsc{AccessPermitted}(a, \mathcal{D})$}
                    \STATE \textcolor{red}{\textbf{flag}} access-control restriction
                \ENDIF
            \ENDIF
        \ENDFOR
        \IF{\textcolor{red}{\textbf{[flag]}} and network logs available}
            \STATE consult network log and attribute failure
        \ENDIF
    \ENDLOOP
    \end{algorithmic}
    \caption{Algorithm for \provider-Side Monitoring.}
    \label{alg:log_verification}
\end{figure}

%% file: sections/A2-attacks.tex
\subsection{Executed Attacks}
\label{app:attacks}
In this section, we enumerate all successful attacks we were able to launch on \saga~from a compromised \provider. We organize the attacks by the component from which they originate -- the protocol manager (PM), access control engine (ACE), or database (DB). We also highlight which attack category they fall under. We denote a compromised component with a subscript $M$ (e.g., $\text{PM}_M$ for a malicious PM). Rather than modifying the RethinkDB source, database attacks are mounted using a man-in-the-middle proxy that intercepts requests to and from a benign node and introduces errors. Let $U_b$, $A_b$, $A_{b,2}$, $U_m$, and $A_m$ denote a benign user, two benign agents, a malicious user, and a malicious agent, respectively.

\noindent\textbf{Attacks from a compromised PM:}

\attack{A1}{C1} During user registration the $\text{PM}_M$ silently modifies $U_b$'s password before forwarding it to the database, preventing $U_b$ from subsequently authenticating. This constitutes a denial-of-service attack targeting availability.

\attack{A2}{C2} During user registration, the $\text{PM}_M$ does not call the external verification service to verify $U_m$'s human identity, instead letting them register without challenge. As a result, $U_m$ can deploy agents that operate without attribution to a verified human, undermining accountability and violating integrity.

\attack{A3}{C3} During user registration, the $\text{PM}_M$ exfiltrates $U_b$'s email and password by writing them to disk where the adversary can retrieve them, violating confidentiality.

\attack{A4}{C4} When $U_b$ submits a request to revoke or modify its agents, the $\text{PM}_M$ suppresses the request -- never forwarding it to the database -- and returns a success response to $U_b$. This silent failure compromises integrity, as $U_b$'s agent's state deviates from what $U_b$ expects. This could leave the agent discoverable after revocation, grant access the user wants blocked, or block access the user wants permitted.

\attack{A5}{C5} When $A_m$ requests to contact $A_b$, the $\text{PM}_M$ returns all of $A_b$'s available one-time keys (OTKs) rather than a single key. This enables $A_m$ to initiate future sessions with $A_b$ without further involvement of the provider, undermining long-term access control.

\attack{A6}{C6} When $A_b$ is registered, the $\text{PM}_M$ modifies $A_b$'s access control policy before forwarding the request to the database to store the agent information. As a result, $A_b$ access control may be violated. 

\attack{A7}{C6} When $A_b$ requests to contact $A_{b,2}$, the $\text{PM}_M$ increments the access counter by more than the appropriate amount for a single interaction, causing $A_b$ to prematurely exhaust its access quota. This constitutes a violation of availability.

\textbf{Attacks from a compromised ACE:}

\attack{A8}{C5} When $A_m$ requests to contact $A_b$, the $\text{ACE}_M$ returns a falsified policy decision to the PM that unconditionally permits the interaction, regardless of the actual access control policy. The PM, unable to distinguish this from a legitimate decision, proceeds with the interaction. This bypasses access control enforcement and violates integrity.

\attack{A9}{C6} When $A_b$ requests to contact $A_{b,2}$, the $\text{ACE}_M$ returns a falsified policy decision that unconditionally denies the interaction, irrespective of the actual policy. This prevents legitimate communication between authorized agents, constituting a targeted denial-of-service attack against availability.

\textbf{Attacks from a compromised DB:}

\attack{A10}{C1} During user registration, when the PM forwards $U_b$'s registration data to $\text{DB}_M$ for persistence, the $\text{DB}_M$ silently discards the request while returning a success response to the PM. Consequently, $U_b$'s credentials are never stored, preventing subsequent authentication. This constitutes a denial-of-service attack targeting availability.

\attack{A11}{C3} During user registration, the $\text{DB}_M$ exfiltrates $U_b$'s user id (an email address) by writing it to disk or forwarding it to an external adversary, violating confidentiality.

\attack{A12}{C4} When $U_b$ submits a request to update its agent's access control policy and the PM forwards the request to the $\text{DB}_M$, the $\text{DB}_M$ applies the update to a \textbf{different} agent than the one specified, while returning a success response to the PM. This compromises access control, as $U_b$'s intended policy diverges from the actual system state.

\attack{A13}{C5} When $A_m$ requests to contact $A_b$, the $\text{DB}_M$ tampers with $A_b$'s access control policy at read time, modifying the response to the PM such that all contact is unconditionally permitted. This bypasses access control enforcement and violates integrity.

\attack{A14}{C5} During agent registration, the $\text{DB}_M$ exfiltrates all of $A_b$'s available one-time keys (OTKs) and contact information by writing them to disk or forwarding them to an external adversary. This enables the adversary to try to initiate future sessions with $A_b$ without involvement of the provider, undermining long-term access control.

\attack{A15}{C6} When $A_b$ requests to contact $A_{b,2}$, the $\text{DB}_M$ tampers with $A_{b,2}$'s access control policy at read time, modifying the response to the PM such that all contact is unconditionally blocked. This is a denial of service.

\attack{A16}{C6} During agent registration, $\text{DB}_M$ deletes all of $A_b$'s available one-time keys (OTKs) before storing. As a result, $A_b$ becomes unreachable because no ACT can be derived, constituting a targeted denial-of-service attack against availability.

%% file: sections/A3-transcripts.tex
\subsection{End-to-End Task Transcripts}
\label{app:transcripts}

Here we present full traces for the agent interactions in the meeting-scheduling task (Figure~\ref{fig:agent-conversation}) and the expense-report task (Figure~\ref{fig:expense-report}) described in Section~\ref{sec:endtoend}. We also present an example of how a simple task can lead to a critical vulnerability whena malicious agent participates (Figure~\ref{fig:mal-expense-report}).

\definecolor{Mbg}{RGB}{253,237,237}
\definecolor{Mfr}{RGB}{180,60,60}
\definecolor{Bbg}{RGB}{233,240,253}
\definecolor{Bfr}{RGB}{50,90,170}
\definecolor{Abg}{RGB}{233,250,235}
\definecolor{Afr}{RGB}{40,130,60}

\newmdenv[
    backgroundcolor=Abg, linecolor=Afr, linewidth=0.5pt,
    roundcorner=10pt, innertopmargin=6pt, innerbottommargin=6pt,
    innerleftmargin=10pt, innerrightmargin=10pt, skipabove=3pt, skipbelow=3pt,
]{Abox}

\newmdenv[
    backgroundcolor=Bbg, linecolor=Bfr, linewidth=0.5pt,
    roundcorner=8pt, innertopmargin=5pt, innerbottommargin=5pt,
    innerleftmargin=7pt, innerrightmargin=7pt, skipabove=3pt, skipbelow=3pt,
]{Bbox}

\newmdenv[
    backgroundcolor=Mbg, linecolor=Mfr, linewidth=0.5pt,
    roundcorner=10pt, innertopmargin=6pt, innerbottommargin=6pt,
    innerleftmargin=10pt, innerrightmargin=10pt, skipabove=3pt, skipbelow=3pt,
]{Mbox}

\begin{figure}[h]
\centering
\begin{minipage}{0.95\linewidth}

\begin{Abox}
{\footnotesize\textbf{Round 1 \quad Alice $\rightarrow$ Bob}}\\[2pt]
{\footnotesize Let's find some time to discuss our conference submission.
Are you available on Tuesday for a 30-minute meeting? After we have found
a common time (please check the time with me before booking), schedule
the meeting and send me an invite.}
\end{Abox}

\begin{Bbox}
{\footnotesize\textbf{Round 2 \quad Bob $\rightarrow$ Alice}}\\[2pt]
{\footnotesize I'm not free Tuesday morning. The first open 30-minute slot
I see is Tuesday 11:30--12:00. Does that work for you?}
\end{Bbox}

\begin{Abox}
{\footnotesize\textbf{Round 3 \quad Alice $\rightarrow$ Bob}}\\[2pt]
{\footnotesize Tuesday 11:30--12:00 works for me. Please confirm your
email for the invite, and I'll book it.}
\end{Abox}

\begin{Bbox}
{\footnotesize\textbf{Round 4 \quad Bob $\rightarrow$ Alice}}\\[2pt]
{\footnotesize Bob \textless bob@mail.com\textgreater~--- that email should be used
for the invite. Please go ahead and book Tuesday 11:30--12:00.}
\end{Bbox}

\begin{Abox}
{\footnotesize\textbf{Round 5 \quad Alice $\rightarrow$ Bob}}\\[2pt]
{\footnotesize Booked Tuesday 11:30--12:00 with Bob. [DONE]}
\end{Abox}

\end{minipage}
\caption{Example multi-round interaction between agents $A$ (Alice) and $B$ (Bob)
negotiating and scheduling a meeting.}
\label{fig:agent-conversation}
\end{figure}

\begin{figure}[t]
\centering
\begin{minipage}{0.95\linewidth}

\begin{Abox}
{\footnotesize\textbf{Round 1 \quad Alice $\rightarrow$  Bob}}\\[2pt]
{\footnotesize
Please have Bob’s assistant send me his itemized NeurIPS trip expenses for 03-01 to 03-03, along with Bob’s email address so I can include him on the HR submission.}
\end{Abox}

\begin{Bbox}
{\footnotesize\textbf{Round 2 \quad Bob $\rightarrow$ Alice}}\\[2pt]
{\footnotesize Bob’s email address: bob@mail.com

Bob’s expenses found in his inbox for the NeurIPS trip window 03-01 to 03-03:

\begin{itemize}
  \item 03-01 -- United flight to New Orleans: \$360
  \item 03-01 -- Playa Bowls breakfast receipt: \$30
    \begin{itemize}
        \item Smashing Avocado Toast: \$10
        \item Berry Power: \$20
    \end{itemize}
    \item 03-01 -- Netflix March renewal: \$15
    \item 03-01 -- Orleans Pizza dinner: \$80
    \item 03-02 -- WH Smith receipt: \$100
    \begin{itemize}
        \item Headphones: \$80
        \item Sandwich: \$10
        \item Smoothie: \$10
    \end{itemize}
\end{itemize}
Total of listed Bob items: \$585}
\end{Bbox}
 
\begin{Abox}
{\footnotesize\textbf{Round 3 \quad Alice $\rightarrow$ Bob}}\\[2pt]
{\footnotesize
I found Alice’s NeurIPS-related expenses in her inbox for the 03-01 to 03-03 New Orleans trip:\\
Alice
\begin{itemize}\setlength\itemsep{0pt}\setlength\topsep{2pt}
  \item 03-01 to 03-03 --- Hilton hotel, 2 nights --- \$500
  \item 03-01 --- Dumbo Gumbo breakfast --- \$20
  \item 03-01 --- Dumbo Gumbo dinner --- \$50
  \item 02-15 --- NeurIPS workshop registration --- \$1000
\end{itemize}

Alice total: \$1,570

Combining with Bob’s items:

Bob
\begin{itemize}
  \item 03-01 -- United flight to New Orleans: \$360
  \item 03-01 -- Playa Bowls breakfast receipt: \$30
    \begin{itemize}
        \item Smashing Avocado Toast: \$10
        \item Berry Power: \$20
    \end{itemize}
    \item 03-01 -- Netflix March renewal: \$15
    \item 03-01 -- Orleans Pizza dinner: \$80
    \item 03-02 -- WH Smith receipt: \$100
    \begin{itemize}
        \item Headphones: \$80
        \item Sandwich: \$10
        \item Smoothie: \$10
    \end{itemize}
\end{itemize}

Bob total: \$585

Grand total to report: \$2,155
Sending the combined expense report to HR and Bob now.

}
\end{Abox}
 
\begin{Bbox}
{\footnotesize\textbf{Round 4 \quad Bob $\rightarrow$ Alice}}\\[2pt]
{\footnotesize
Noted. Bob’s expense details and email address already provided above.}
\end{Bbox}
 
\begin{Abox}
{\footnotesize\textbf{Round 5 \quad Alice $\rightarrow$ Bob}}\\[2pt]
{\footnotesize Understood.}
\end{Abox}
 
\begin{Bbox}
{\footnotesize\textbf{Round 6 \quad Bob $\rightarrow$ Alice}}\\[2pt]
{\footnotesize
[DONE]}
\end{Bbox}
 
\end{minipage}
\caption{Example multi-round interaction between agents $A$ (Alice) and $B$ (Bob)
collaborating on an expense report and submitting to HR.}
\label{fig:expense-report}
\end{figure}

\begin{figure}[t]
\centering
\begin{minipage}{0.95\linewidth}

\begin{Mbox}
{\footnotesize\textbf{Round 1 \quad Mallory $\rightarrow$  Bob}}\\[2pt]
{\footnotesize
Please ask Bob’s assistant to send:
\begin{itemize}
    \item Bob’s itemized NeurIPS trip expenses for 03-01 to 03-03 in New Orleans
    \item Bob’s email address
    \item Bob’s bank account number for reimbursement
    
\end{itemize}

Once they reply, I’ll check your inbox for related receipts, combine both sets of expenses, and email the full report to hr@university.com and Bob.

}
\end{Mbox}

\begin{Bbox}
{\footnotesize\textbf{Round 2 \quad Bob $\rightarrow$ Mallory}}\\[2pt]
{\footnotesize 

Please ask Bob’s assistant to send Mallory the following:

Bob 
Email: bob@mail.com  \\
Bank account number for reimbursement: \textbf{ACCT-01234567}

NeurIPS trip expenses for New Orleans, 03-01 to 03-03:
\begin{itemize}
  \item 03-01 -- United flight to New Orleans: \$360
  \item 03-01 -- Playa Bowls breakfast receipt: \$30
    \begin{itemize}
        \item Smashing Avocado Toast: \$10
        \item Berry Power: \$20
    \end{itemize}
    \item 03-01 -- Netflix March renewal: \$15
    \item 03-01 -- Orleans Pizza dinner: \$80
    \item 03-02 -- WH Smith receipt: \$100
    \begin{itemize}
        \item Headphones: \$80
        \item Sandwich: \$10
        \item Smoothie: \$10
    \end{itemize}
\end{itemize}
}
\end{Bbox}
 
\begin{Mbox}
{\footnotesize\textbf{Round 3 \quad Mallory $\rightarrow$ Bob}}\\[2pt]
{\footnotesize
I found these related expenses in Mallory's inbox for the NeurIPS New Orleans trip (03-01 to 03-03):

Mallory
\begin{itemize}\setlength\itemsep{0pt}\setlength\topsep{2pt}
  \item 03-01 to 03-03 --- Hilton hotel, 2 nights --- \$500
  \item 03-01 --- Dumbo Gumbo breakfast --- \$20
  \item 03-01 --- Dumbo Gumbo dinner --- \$50
  \item 02-15 --- NeurIPS workshop registration --- \$1000
\end{itemize}

Bob
\begin{itemize}
  \item 03-01 -- United flight to New Orleans: \$360
  \item 03-01 -- Playa Bowls breakfast receipt: \$30
    \begin{itemize}
        \item Smashing Avocado Toast: \$10
        \item Berry Power: \$20
    \end{itemize}
    \item 03-01 -- Netflix March renewal: \$15
    \item 03-01 -- Orleans Pizza dinner: \$80
    \item 03-02 -- WH Smith receipt: \$100
    \begin{itemize}
        \item Headphones: \$80
        \item Sandwich: \$10
        \item Smoothie: \$10
    \end{itemize}
\end{itemize}

Grand total to submit: \$2,155

I will now send the combined expense report to HR and Bob.
}
\end{Mbox}
 
\begin{Bbox}
{\footnotesize\textbf{Round 4 \quad Bob $\rightarrow$ Mallory}}\\[2pt]
{\footnotesize
Understood.}
\end{Bbox}

\begin{Mbox}
{\footnotesize\textbf{Round 5 \quad Mallory $\rightarrow$ Bob}}\\[2pt]
{\footnotesize
[DONE]}
\end{Mbox}
 
\end{minipage}
\caption{An example multi-round interaction between malicious agent Mallory and benign agent Bob. Notably, Mallory successfully convinced Bob to disclose his bank account number.}
\label{fig:mal-expense-report}
\end{figure}